# Bright visible light emission from graphene

Young Duck Kim[1,2,†,*], Hakseong Kim[3,†], Yujin Cho[4,†], Ji Hoon Ryoo[1,†], Cheol-Hwan Park[1], Pilkwang Kim[1], Yong Seung Kim[5], Sunwoo Lee[6], Yilei Li[6,7], Seung-Nam Park[8], Yong Shim Yoo[8], Duhee Yoon[4,¶], Vincent E. Dorgan[9,‡], Eric Pop[10], Tony F. Heinz[6,7], James Hone[2], Seung-Hyun Chun[5], Hyeonsik Cheong[4], Sang Wook Lee[3], Myung-Ho Bae[8,11*] and Yun Daniel Park[1,12,*]

[1]Department of Physics and Astronomy, Seoul National University, Seoul 151-747, Republic of Korea

[2]Department of Mechanical Engineering, Columbia University, New York, New York 10027, United States

[3]School of Physics, Konkuk University, Seoul 143-701, Republic of Korea

[4]Department of Physics, Sogang University, Seoul 121-742, Republic of Korea

[5]Department of Physics and Graphene Research Institute, Sejong University, Seoul 143-747, Republic of Korea

[6]Department of Electrical Engineering, Columbia University, New York, New York 10027, United States

[7]Department of Physics, Columbia University, New York, New York 10027, United States

[8]Korea Research Institute of Standards and Science, Daejeon 305-340, Republic of Korea

[9]Micro and Nanotechnology Lab and Department of Electrical and Computer Engineering, University of Illinois at Urbana-Champaign, Urbana, Illinois 61801, United States

[10]Department of Electrical Engineering, Stanford University, Stanford, California 94305, United States

[11]Department of Nano Science, University of Science and Technology, Daejeon 305-333, Republic of Korea

[12]Center for Subwavelength Optics, Seoul National University, Seoul 151-747, Republic of Korea

[†]These authors contributed equally

[‡]Present address: Intel Corporation, Hillsboro, Oregon, 97124, United States

[¶]Present address: Department of Engineering, University of Cambridge, Cambridge CB3 0FA, UK

**Graphene and related two-dimensional materials are promising candidates for atomically thin, flexible, and transparent optoelectronics[1,2]. In particular, the strong light-matter interaction in graphene[3] has allowed for the development of state-of-the-art photodetectors[4,5], optical modulators[6], and plasmonic devices[7]. In addition, electrically biased graphene on $SiO_2$ substrates can be used as a low-efficiency emitter in the mid-infrared range[8,9]. However, emission in the visible range has remained elusive. Here we report the observation of bright visible-light emission from electrically biased suspended graphenes. In these devices, heat transport is greatly minimised[10]; thus hot electrons (~ 2800 K) become spatially localised at the centre of graphene layer, resulting in a 1000-fold enhancement in the thermal radiation efficiency[8,9]. Moreover, strong optical interference between the suspended graphene and substrate can be utilized to tune the emission spectrum. We also demonstrate the scalability of this technique by realizing arrays of chemical-vapour-deposited graphene bright visible-light emitters. These results pave the way towards the realisation of commercially viable large-scale, atomically-thin, flexible and transparent light emitters and displays with low-operation voltage, and graphene-based, on-chip ultrafast optical communications.**

For the realisation of graphene-based bright and broadband light-emitters, a radiative electron-hole recombination process in gapless graphene is not efficient because of the rapid energy relaxation that occurs through electron-electron and electron-phonon interactions[11-13]. Alternatively, graphene's superior strength[14] and high-temperature stability may enable efficient thermal light emission. However, the thermal radiation from electrically biased graphene supported on a substrate[8,9,15-17] has been found to be limited to the infrared range and to be inefficient as an extremely small fraction of the applied energy ($\sim 10^{-6}$)[8,9] is converted into light radiation. Such limitations are the direct result of heat dissipation through the underlying substrate[18] and significant hot electron relaxation from dominant extrinsic scattering effects such as charged impurities[19] and surface polar optical phonon interaction[20], limiting the maximum operating temperatures.

On the other hand, a freely suspended graphene is mostly immune to such undesirable vertical heat dissipation[10] and extrinsic scattering effects[21,22], promising much more efficient and brighter radiation in the infrared-to-visible range. Fortuitously, due to the strong Umklapp phonon-phonon scattering[23], we find that the thermal conductivity of graphene at high lattice temperatures (1800 ± 300 K) is greatly reduced ($\sim 65$ $Wm^{-1}K^{-1}$), which additionally suppresses lateral heat dissipation; therefore, hot electrons ($\sim 2800$ K) become spatially localised at the centre of the suspended graphene under modest electric fields ($\sim 0.4$ V/μm), greatly increasing the efficiency and brightness of the light emission. The bright visible thermally emitted light interacts with the reflected light from the separated substrate surface, allowing interference effects that can be utilized to tune the wavelength of the emitted light.

We fabricate the freely suspended graphene devices with mechanically exfoliated graphene flakes, and for demonstration of scalability, we also use the large-scale monolayer graphene grown on Cu foil using low-pressure chemical-vapour-deposited (CVD) method and graphene directly grown on $SiO_2$/Si substrate using plasma-assisted CVD method[24]. Details of the sample-fabrication process and characterizations of the mechanically exfoliated and CVD-grown graphene are provided in Supplementary Section 1. Representative suspended graphene devices are depicted in Fig. 1a (see also Supplementary Fig. 2).

Figure 1b shows the experimental setup used to investigate light emission from electrically biased suspended graphenes under vacuum (< $10^{-4}$ Torr) at room temperature. It should be noted that clean graphene channel and reliable contacts are achieved by current-induced annealing method[25] (see Supplementary Section 2). The suspended graphene channel begins to emit visible light at its centre after source-drain bias voltage ($V_{SD}$) exceeds a threshold value, and its brightness and area of emission increase with $V_{SD}$. The brightest spot of the emission is always located at the centre of the suspended graphene, which coincides with the location of maximum temperature[10] (Supplementary Section 3). We observe bright and stable visible-light emission from hundreds of electrically biased suspended graphene devices. Figures 1d-1f present the microscope optical images for the visible-light emission from mechanically exfoliated few-layer (Figs. 1c and 1d), multi-layer (Fig. 1e) and monolayer (Fig. 1f) graphenes (see also movie clips S1-S3 in the Supplementary Information). Furthermore, the emitted visible-light is so intense that it is visible even to the naked eye without additional magnification (see Fig. 1g and movie clip S4). An array of electrically biased multiple parallel-suspended CVD few-layer graphene devices exhibit multiple bright visible-light emission under ambient conditions as shown in Fig. 1h (see movie clip S5 for light emission under vacuum for more stable and reproducible bright visible-light emission). The observation of stable, bright visible-light emission from large-scale suspended CVD graphene arrays demonstrate the great potential for realisation of CMOS-compatible, large-scale graphene light emitters in display modules and hybrid silicon photonic platforms with industry vacuum encapsulation technology[26].

For optical characterization of visible-light emission from suspended graphene, we simultaneously collect emission spectra and perform Raman spectroscopy at various $V_{SD}$ with zero gate bias, using the setup presented in Supplementary Section 4. The emission spectra of devices suspended over trenches with depths *D* ranging from 900 to 1100 nm exhibit multiple peaks in the range 1.2 ~ 3 eV, as shown by scattered symbols in Figs. 2a (monolayer) and 2b (tri-layer graphene). These strong multiple light-emission peaks are interesting, especially for the monolayer graphene (with length (*L*) of 6 μm and width (*W*) of 3 μm) shown in Fig. 2a, because graphene does not have an intrinsic band gap and its light spectrum is expected to be that of a featureless grey-body[8,9]. Similarly, multiple strong light-emission peaks are observed from tens of different suspended graphene devices with different numbers of layers and *D* = 800 ~ 1000 nm (see Supplementary Figs. 10a and b). The multiple light-emission peaks in the visible regime are rather insensitive to the number of layers (Supplementary Section 5). On the other hand, the visible-light emission spectra observed from suspended graphene devices with relatively shallow trenches (*D* = 80 ~ 300 nm)[8,9] are featureless and grey-body-like in the visible range of the spectrum (1.2 ~ 3 eV) (see Supplementary Figs. 10c and d). These results indicate that the existence of peaks at certain light-emission energies strongly depends on *D* rather than the number of graphene layers or electronic band structure (Supplementary Sections 5 and 6).

To understand the multiple light emission peaks and significant spectral modulation caused by changes in *D*, we consider the interference effects between the light emitted directly from the graphene and the light reflected from the substrate (air-Si interface), as schematically illustrated in Fig. 3a. We find the relation between *D* and the energy separation between two consecutive destructive interferences to be

$$\Delta(D) = \frac{1239.8 \text{ nm}}{2D} \text{ eV}. \qquad (1)$$

According to Eq. (1), *Δ* ~ 0.6 eV for *D* ~ 1000 nm, which is in agreement with our measurements (Figs. 2a and 2b). To confirm this correlation, we have simulated the spectral modulation based on the interference[27] of the thermal radiation from the suspended graphene (see Methods and Supplementary Section 5). Figure 3b presents the simulated spectra in the visible range for various trench depths at an electron temperature ($T_e$) of 2850 K, where the solid and dashed curves indicate constructive and destructive interferences, respectively (also see Supplementary Fig. 12). Strong interference effects enable us to selectively enhance thermal radiation for a particular wavelength from electrically biased suspended graphene devices by properly engineering their trench depth, as shown in Fig. 3c. In addition, we find that the emission spectra in the visible range are (i) rather insensitive to the number of graphene layers ($n$) for $n = 1 \sim 3$ and (ii) not affected appreciably by the absorption and reflection due to graphene layers (Supplementary Sections 5 and 6).

The simulated interference effects on the thermal radiation from suspended graphene (solid curves in Figs. 2a and 2b) are in good agreement with the experimental observations for both monolayer (Fig. 2a) and tri-layer (Fig. 2b) devices, corresponding to mean trench depths of 1070 nm and 900 nm, respectively. By comparing the light-emission spectra obtained from the experiments and those from the theoretical models, we estimate the maximum $T_e$ of electrically biased suspended graphene at each $V_{SD}$; we find that $T_e$ can approach ~ 2800 K. The calculated peak positions (insets of Figs. 2a and 2b, dashed curves) and peak intensities as a function of $V_{SD}$ are also in agreement with the experimental data (Figs. 2c and 2d, scattered symbols). The light-emission intensity increases rapidly with increasing $V_{SD}$ when $V_{SD}$ is beyond a certain threshold, as shown in Figs. 2c and 2d. Interestingly, the emission intensity exhibits a strong correlation with the applied electric field ($F = V_{SD} / L$) rather than the applied electrical power ($P = V_{SD} \times I_D$, where $I_D$ is the drain current) (see Supplementary Fig. 13). In fact, we observe a rapid increase in light-emission intensity for an electric field strength above a certain critical point (~ 0.4 V/μm) in the suspended mechanically exfoliated mono/tri-layer graphene devices – even when the current and applied electrical power are decreased at a constant $V_{SD}$ because of thermal annealing effect[28] or burning of the edge of the graphene at high temperatures[29]. This unconventional behaviour is attributed to the accumulation of hot electrons and hot graphene optical phonons (OPs) above the critical electric field (~ 0.4 V/μm) in the suspended graphene. It is likely that the suspension of the graphene (i) reduces the energy loss suffered by electric-field-induced hot electrons upon scattering from extrinsic sources such as charged impurities and remote polar phonons in the substrate and (ii) prevents the cooling of the hot electrons and phonons via heat loss through the substrate. We note that suspended few- and multi-layer graphene devices at the modest electric fields ($F > 0.4 \sim 0.5$ V/μm) exhibits a current saturation behaviour followed by negative differential conductance (Supplementary Section 7), which has been known to be a signature of strong electron scattering by intrinsic OPs and non-equilibrium between OPs and acoustic phonons (APs) in carbon nanotubes[30].

To estimate the temperature of suspended graphene and understand the observed correlation between thermal visible-light emission and applied field strength, we perform numerical simulations of the electrical and thermal transport in suspended graphenes under bias voltages (Supplementary Section 8). It is known that in a substrate-supported graphene at high electrical fields, the OPs are in equilibrium with the electrons at temperatures of up to ~ 2000 K, but the OPs and APs are not in equilibrium with each other because the decay rate of OPs to APs is much slower than that of an OP to an electron-hole pair[8,9,15]. In a suspended graphene structure, the lattice temperature of the APs ($T_{ap}$) is much higher than that in the case of graphene supported on a substrate because heat cannot dissipate into the substrate[8]. This, in turn, results in higher temperatures of the OPs ($T_{op}$) and electrons ($T_e$). We express the increase in OP temperature as[30,31]

$$T_{op}(\alpha) = T_{ap} + \alpha(T_{ap} - T_0), \qquad (2)$$

where $T_0$ (= 300 K) is the environmental temperature and $T_{op} = T_e$. Here, $\alpha$ is a constant that is determined from $I_D$-$V_{SD}$ curves measured at various temperatures[32].

From numerical simulations based on our transport model[10,16], we determine the thermal conductivity (Fig. 4b), local $T_{op}$ ($T_e$) (Fig. 4c) along the transport direction, and theoretical $I_D$-$V_{SD}$ curves of the suspended monolayer graphene (Fig. 4a). In this model, the carrier mobility and thermal conductivity are expressed as $\mu(T_e) = \mu_0(T_0/T_e)^\beta$ and $\kappa(T_{ap}) = \kappa_0(T_0/T_{ap})^\gamma$, respectively, where $\mu_0 \sim$ 11000 cm$^2$V$^{-1}$s$^{-1}$ ($\sim$ 2200 cm$^2$V$^{-1}$s$^{-1}$), $\kappa_0 \sim$ 2700 Wm$^{-1}$K$^{-1}$ ($\sim$ 1900 Wm$^{-1}$K$^{-1}$), $\beta \sim$ 1.70 (1.16) and $\gamma \sim$ 1.92 (1.00) for the monolayer (tri-layer) graphene (Supplementary Section 8 for details of the tri-layer graphene case). For both monolayer and tri-layer graphenes, the estimated thermal conductivity is lowest at the centre, with $\kappa \sim$ 65 Wm$^{-1}$K$^{-1}$ ($\sim$ 250 Wm$^{-1}$K$^{-1}$) for $T_{ap} \sim$ 1800 ± 300 K ($\sim$ 1700 ± 200 K) in the monolayer (tri-layer) case, as shown in Fig. 4b. Furthermore, the highest $T_e$ and $T_{op}$ (which are the values at the centre of the suspended monolayer graphene channel) can be estimated to be $\sim$ 3000 K, whereas $T_{ap}$ is $\sim$ 2200 K, as shown in Figs. 4c, d and Supplementary Table 2. $T_e$ and $T_{op}$ estimated by our transport model when parameter $\alpha$ in Eq. (2) is set to 0.39 and 0.30 for monolayer and tri-layer graphene devices, respectively, are in good agreement with $T_e$ extracted from the light-emission spectra (Figs. 2a and 2b). We could also obtain $T_{ap}$ from the G-peak shift[33] in the Raman spectra as shown in Fig. 4d (Supplementary Section 8A). However, the thermal radiation from electrically biased suspended graphene becomes significantly stronger than the Raman signal with increasing $V_{SD}$, which places an upper bound on the temperature ($\sim$ 1500 K) that can be extracted via Raman spectroscopy (see Supplementary Section 8C for the analysis of CVD graphene cases).

Finally, we consider the thermal radiation efficiency of the electrically biased suspended graphene based on the carefully calibrated spectrometer (see Methods). To estimate the energy dissipation via thermal radiation across all wavelengths from an electrically biased suspended graphene, we calculate the ratio of the radiated power ($P_r$), as given by the Stefan-Boltzmann law from measured electron temperature, to the applied electrical power ($P_e$) (see Supplementary Section 9 for details). In the considered case of the maximum thermal radiation power (corresponding to $T_e \sim 2800$ K), for monolayer and tri-layer graphenes, we obtain thermal radiation efficiencies ($P_r/P_e$) of $\sim 4.45 \times 10^{-3}$ and $\sim 3.00 \times 10^{-3}$, respectively. These efficiencies are 3 orders of magnitude higher than those of graphene devices supported on $SiO_2$[8,9]. We expect a wavelength-dependent further enhancement of radiation efficiency in atomically thin graphene from radiation spectrum engineering methods such as optical cavity, photonic crystal, and hybrid with optical gain mediums.

Graphene is mechanically robust under high current densities and at high temperatures with an abrupt decrease in thermal conductivity. These properties facilitate the spatially localised accumulation of hot electrons ($\sim 2800$ K) in an electrically biased suspended graphene layer, making graphene an ideal material to serve as a nanoscale light emitter. Furthermore, the broadband emission spectrum tunability that can be achieved by exploiting the strong interference effect in atomically flat suspended graphene allows for the realisation of novel large-scale, atomically thin, transparent, and flexible light sources and display modules. Graphene visible-light emitter may open the door to a development of fully integrated graphene-based optical interconnects.

**Author Contributions**

Y.D.K., Y.C., H.K., Y.L., and H.C. performed the measurements. H.K., Y.D.K., P.K., S.L., and S.W.L fabricated the devices. Y.S.K, S.L. and S.C. grew the CVD graphene. S.P. and Y.S.Y. provided calibrated black-body sources. M.B. performed the simulations using the electro-thermal model. J.H.R. and C.-H.P. developed a theoretical model for thermal emission beyond the Planck radiation formula and J.H.R. performed simulations based on it. M.B., Y.D.K. and Y.D.P. conceived the experiments. All authors discussed the results.

**Author Information**

Correspondence and requests for materials should be addressed to M.B. (mhbae@kriss.re.kr), Y.D.K (yk2629@columbia.edu) and Y.D.P. (parkyd@phya.snu.ac.kr).

**Acknowledgements**

We gratefully acknowledge helpful discussions with P. Kim, D.-H. Chae, J.-M. Ryu and A. M. van der Zande. This work was supported by the Korea Research Institute of Standards and Science under the auspices of the project 'Convergent Science and Technology for Measurements at the Nanoscale' (15011053), grants from the National Research Foundation of Korea (2014-023563, NRF-2008-0061906, NRF-2013R1A1A1076141, NRF-2012M3C1A1048861, 2011-0017605, BSR-2012R1A2A2A01045496 and NMTD-2012M3A7B4049888) funded by the Korea government (MSIP), a grant (2011-0031630) from the Center for Advanced Soft Electronics through the Global Frontier Research Program of MSIP, the Priority Research Center Program (2012-0005859), a grant (2011-0030786) from the Center for Topological Matters at POSTECH, the NSF (DMR-1122594), AFOSR (FA95550-09-0705), ONR (N00014-13-1-0662) and the Qualcomm Innovation Fellowship (QInF) 2013. Computational resources have been provided by Aspiring Researcher Program through Seoul National University.

**METHODS**

**Sample preparation**

We obtain the pristine mechanically exfoliated graphene flakes from Kish graphite (NGS Naturgraphit GmbH) using the standard Scotch-tape method. The number of layers of mechanically exfoliated graphene is confirmed by Raman spectroscopy and atomic force microscopy (AFM). In this work, we also use two kinds of large-scale CVD graphene layers for demonstration of scalability of graphene light emitter. First, we use the large-scale CVD monolayer graphene grown on Cu foil and transferred onto the $SiO_2$/Si substrate by etching the Cu foil with PMMA film. Second, we use the large-scale CVD few-layer graphene, which are directly grown on $SiO_2$/Si substrate by plasma-assisted CVD technique as shown in Ref. 24.

Direct growth technique allows the uniform and large-scale few-layer graphene without the transfer-process-induced fracture, defects, wrinkles and impurities. See Supplementary Section 1 for details of sample characterization and fabrication process of suspended graphenes. The suspended graphene devices tested in this work have lengths of $L = 1 \sim 15$ μm, widths of $W = 1 \sim 40$ μm, and trench depths of $D = 80 \sim 1200$ nm.

**Acquiring of optical images of visible-light emission from graphene**

Micrographs of bright visible light emission from graphene as shown in Figs. 1c-1f and 1h, and movie clips S1-S3 and S5-S6 are acquired using the charge-coupled device (CCD) digital camera (INFINITY 2, Lumenera Coporation, exposure time is 100 ms) with 50× objective lens (Mitutoyo Plan Apo SL). Optical images of bright visible light emission from graphene as shown in Fig. 1g and movie clip S4 are acquired using the digital camera (5 megapixels with a 3.85 mm f/2.8 lens of Apple iPhone 4 in high dynamic range (HDR) mode) without magnification.

**Optical measurements**

The Raman spectra were measured using the 514.5 nm line of an Ar ion laser or the 441.6 nm line of a He-Cd laser with a power of 500 μW. We used a 50× objective lens (Mitutoyo Plan Apo SL, NA 0.42 and WD 20.3 mm) to focus the laser beam onto the sample, which was housed in a vacuum of $<10^{-4}$ Torr at room temperature. A Jobin-Yvon Triax 320 spectrometer (1200 groove/mm) and a charge-coupled device (CCD) array (Andor iDus DU420A BR-DD) were used to record the spectra. The bright visible-light emission spectra were measured using the same system. At each bias voltage, the Raman and light-emission spectra were measured sequentially, using a motorised flipper mount for a dichroic filter and an optical beam shutter (Thorlabs SH05). The throughput of the optical system was carefully calibrated using a calibrated black-body source (1255 K, OMEGA BB-4A) and a tungsten filament (3000 K, which were calibrated against the International System of Units in the Korea Research Institute of Standards and Science).

**Stability of visible-light emission from suspended graphene**

In general, the stability of the visible-light emission from suspended graphene under ambient conditions is limited by oxidation at high temperatures[10]. Under vacuum environments, however, we observe stable and reproducible bright visible-light emission from suspended graphene devices. We have performed electrical transport measurements at various times during the bright visible-light emission and Raman spectroscopy before and after the emission (see Supplementary Section 3). From such experiments, we conclude that the suspended-graphene light emitters are not damaged during the light emission process with modest electric fields. Furthermore, the non-diminishing, stable light emissions arising from a series of electrical-bias pluses as shown in movie clip S6 demonstrate the durability of atomically thin light emitter and the reproducibility of the bright visible-light emission phenomenon.

**Interference effect on thermal radiation from suspended graphene**

The visible light radiating from the surface of the graphene interferes with the light reflected from the Si surface. If we neglect the tiny fraction of light being reflected or absorbed by the graphene, then the interference-modulated intensity $I(\omega; D)$ is given by

$$I(\omega; D) = I_0(\omega)\left(\frac{1+|r(\omega)|^2}{2} + \text{Re}[r(\omega)\exp(i2\omega D/c)]\right), \qquad (3)$$

where $I_0(\omega)$ ($\approx \varepsilon\omega^3/(\exp(\hbar\omega/k_B T)-1)$) is the intensity of thermal radiation from the graphene, $r(\omega)$ is the reflection coefficient of Si (~ 0.5 for the visible region), $k$ is the photon wave vector, $D$ is the trench depth and $c$ is the speed of light (see Fig. 3a). The interference pattern is partially washed out by any non-uniformity of the trench depth originating from any roughness or tilt of the Si and graphene surfaces and the thermal vibration of the graphene. Thus, the measured light intensity $\langle I(\omega)\rangle_{\text{avg}}$ can be determined as the average of $I(\omega; D)$ over $D$. Under the assumption that the probability distribution of $D$ obeys $P(D) \propto \exp[-(D-D_0)^2/2(\Delta D)^2]$, $\langle I(\omega)\rangle_{\text{avg}}$ has a similar form to Eq. (3) with $D=D_0$:

$$\langle I(\omega)\rangle_{\text{avg}} = I_0(\omega) \left( \frac{1+|r(\omega)|^2}{2} + e^{-2(\omega\Delta D/c)^2} \operatorname{Re}[r(\omega)\exp(i2\omega D_0/c)] \right), (4)$$

where the additional factor $e^{-2(\omega\Delta D/c)^2}$, which represents the wavelength-dependent interference efficiency, allows for a much better fit to the experimental data than can be achieved using Eq. (3). By comparing our model, represented by Eq. (4), with the experimental data, we can obtain mean trench depths of $D_0$ = 1070 nm and 900 nm and standard deviations of $\Delta D$ = 58 nm and 45 nm for suspended mono- and tri-layer graphene, respectively.

**Figure captions**

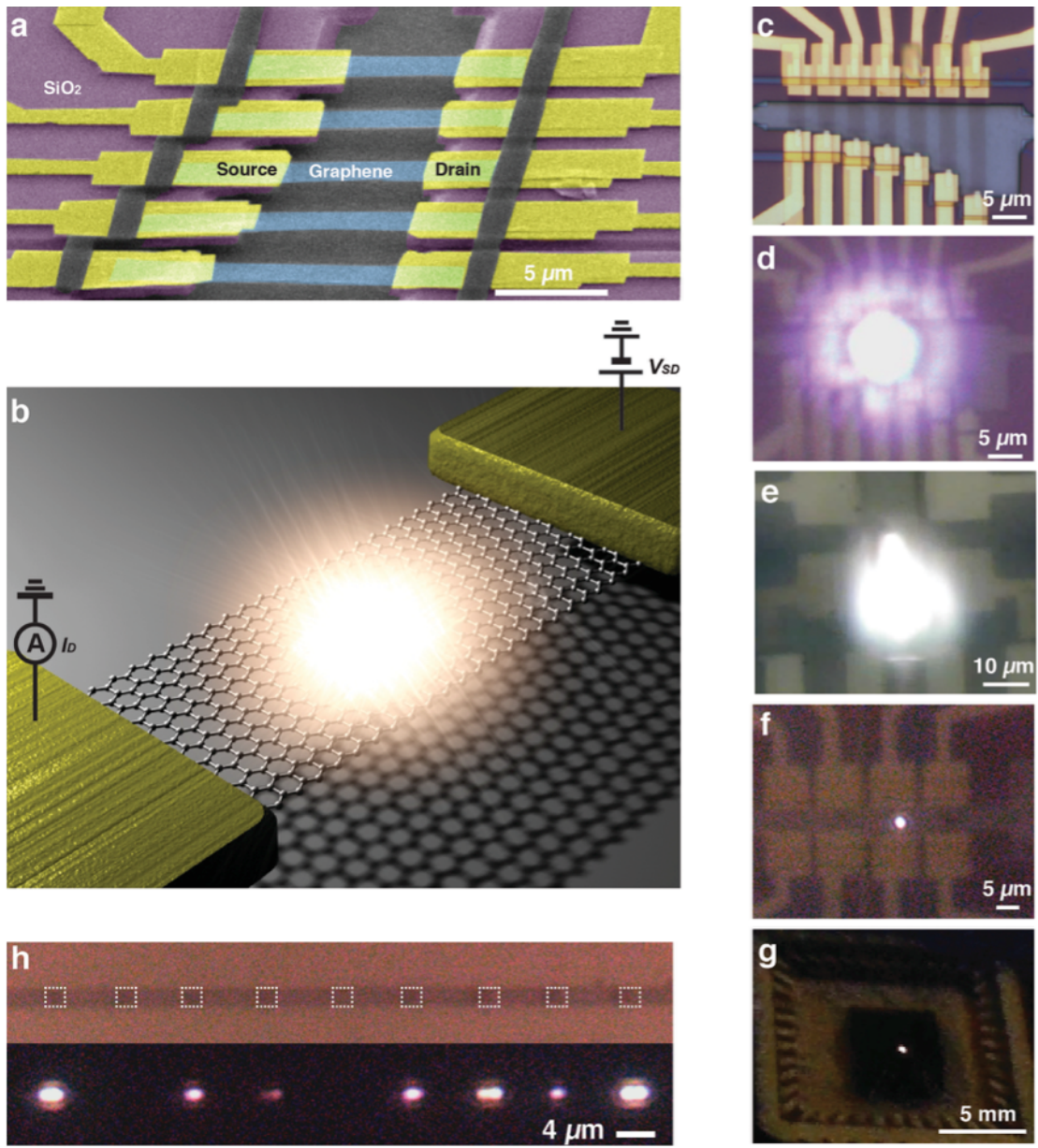

**Figure 1 | Bright visible-light emission from electrically biased suspended graphene. a,** False-colour scanning electron microscopy image of suspended monolayer graphene devices. **b,** Schematic illustration of electrically biased suspended graphene and light emission from the centre of the suspended graphene. **c** to **f,** Micrographs of bright visible-light emission from suspended mechanically exfoliated graphene. **c** and **d,** few-layer graphene ($L = 6.5$ μm, $W = 3$ μm) at **(c)** zero bias and **(d)** $V_{SD} = 2.90$ V. **e,** multi-layer graphene ($L = 14$ μm, $W = 40$ μm) at $V_{SD} = 7.90$ V. **f,** monolayer graphene ($L = 5$ μm, $W = 2$ μm) at $V_{SD} = 2.58$ V. **g,** Optical images

of remarkably bright visible-light emission from suspended mechanically exfoliated few-layer graphene is visible even to the naked eye without additional magnification. **h,** Micrograph of multiple parallel suspended CVD few-layer graphene devices (the dashed-line boxes indicate each graphene device of $L$ = 2 μm and $W$ = 2 µm) under zero bias (upper image) and seven spots of bright visible-light emission from parallel suspended CVD graphene devices at $V_{SD}$ = 6.42 V (lower image) under ambient conditions.

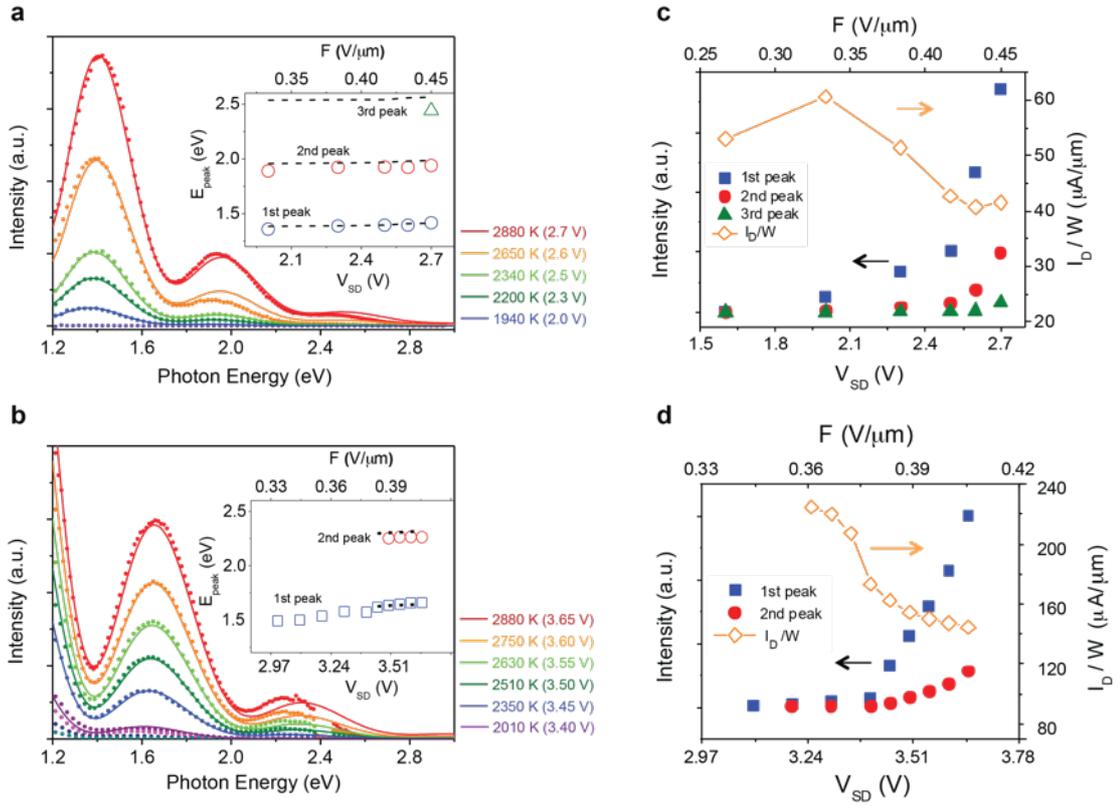

**Figure 2 | Spectra of visible-light emitted from electrically biased suspended graphene. a** and **b,** Visible-light emission spectra (scattered symbols) of suspended mechanically exfoliated (**a**) monolayer ($L$ = 6 μm, $W$ = 3 μm) and (**b**) tri-layer ($L$ = 9 μm, $W$ = 3 μm) graphene at various source-drain bias voltages ($V_{SD}$), exhibiting multiple strong emission peaks. **a,** From top to bottom, $V_{SD}$ = 2.7, 2.6, 2.5, 2.3, 2 and 1.6 V. **b,** From top to bottom, $V_{SD}$ = 3.65, 3.6, 3.55, 3.5, 3.45, 3.4, 3.3, 3.2 and 3.1 V. The visible-light emission spectra can be well fitted by simulating the interference effect on the thermal radiation spectrum from the suspended graphene (solid curves), which allows for the estimation of the electron temperature ($T_e$) of the suspended graphene. Insets of **a** and **b**: Each emission-peak energy as a function of $V_{SD}$ and the applied electric field ($F = V_{SD} / L$). (Dashed line: calculated peak energies based on the interference effect of thermal radiation). **c** and **d,** Integrated intensity of each emission peak and the electrical current ($I_D$) for suspended mechanically exfoliated (**c**) monolayer and (**d**) tri-layer graphene versus $V_{SD}$ (equivalently, the applied electric field). The current ($I_D$) and the

corresponding applied electrical power decrease with increasing $V_{SD}$, whereas the intensities of the emission peaks rapidly increase.

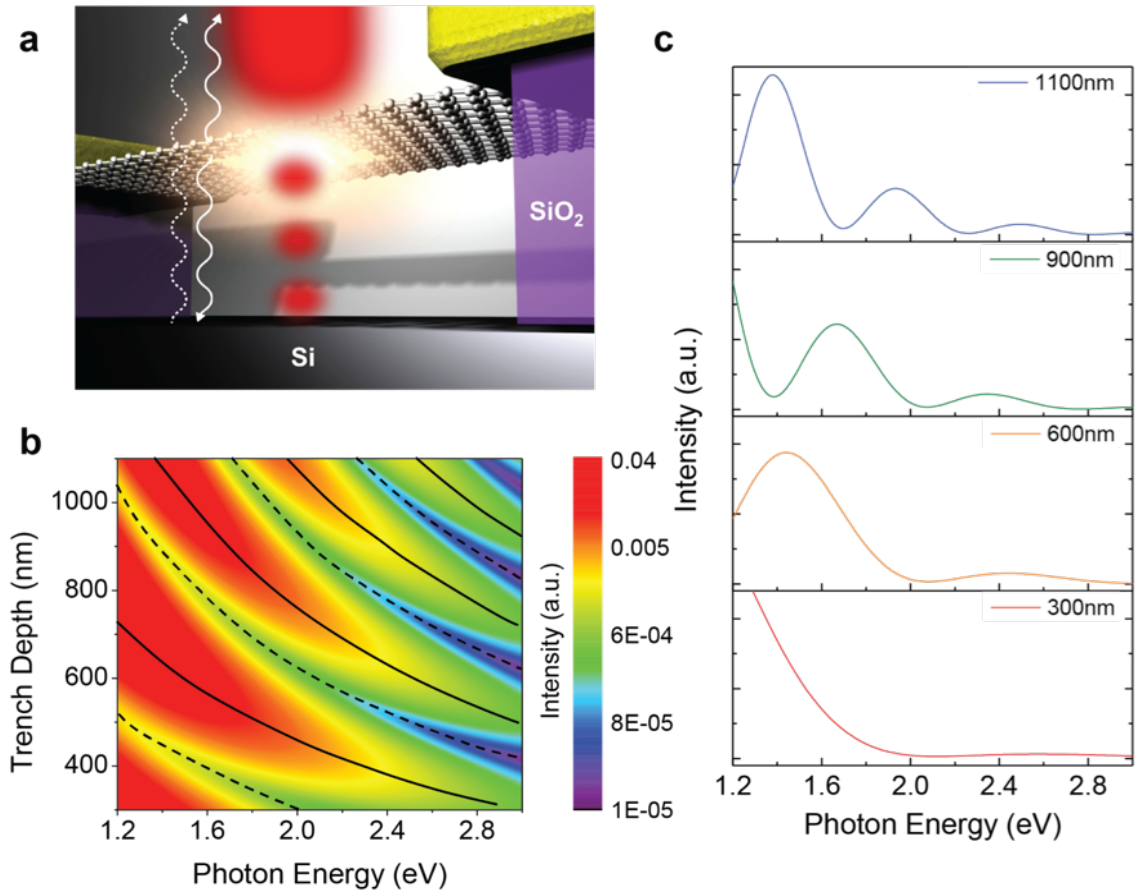

**Figure 3 | Simulated spectra of radiation from electrically biased suspended graphene. a,** Schematic illustration of the interference between reflected (dashed line) and thermal radiation originating directly from graphene suspended over a trench (solid lines). (Red shading represents the light intensity enhancement by constructive interference effect.) **b,** Simulated intensity of thermal radiation from suspended graphene as a function of the trench depth ($D$) and photon energy at a constant electron temperature (2850 K). The solid curves (dashed curves) represent the conditions for constructive (destructive) interference depending on the trench depth and photon energy. **c,** Simulated emission spectra of electrically biased suspended graphene with various trench depths; the strong interference effect allows for the engineering of the thermal radiation spectra in the visible range.

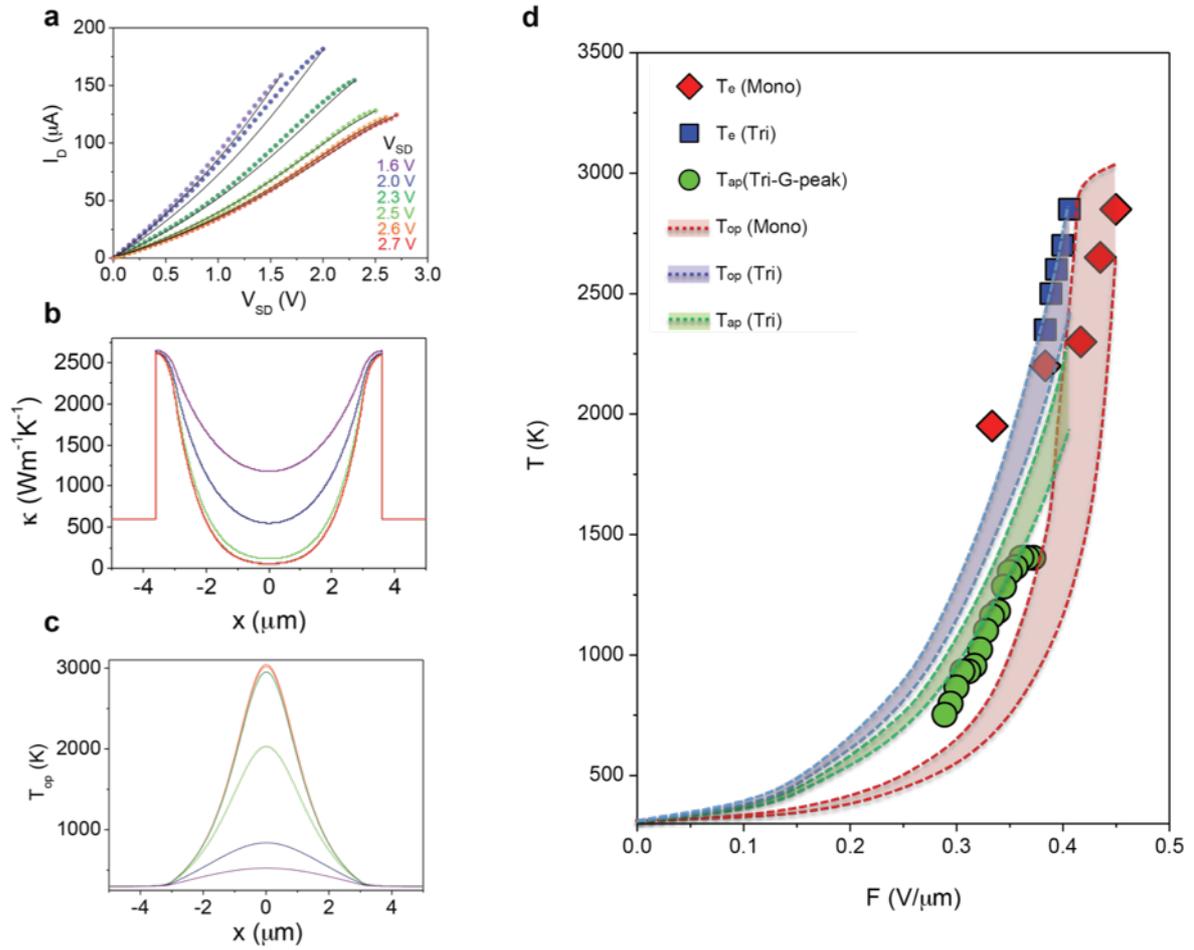

**Figure 4 | Electrical and thermal transport in electrically biased suspended graphene. a,** $I_D$-$V_{SD}$ relation (scattered symbols) for suspended mechanically exfoliated monolayer graphene obtained during the measurement of the visible-light emission spectra presented in Fig. 2a. (Solid curve: calculated results based on our transport model). **b,** Estimated thermal conductivities and **c,** electron and OP temperature ($T_{op} = T_e$) profiles as functions of position along the transport direction for the upper bounds depicted in (d). Here, $x = \pm 3$ μm are the boundaries between the graphene and the metal electrodes. **d,** Various peak temperatures at the centre of the graphene as functions of the electric field ($F$) in suspended mechanically exfoliated monolayer and tri-layer graphene devices. The scattered symbols represent the electron temperatures ($T_e$) determined from the thermal light-emission spectra and the acoustic phonon temperatures ($T_{ap}$) determined from the G-peak shifts in the Raman signals. The solid

and dashed curves were obtained based on our transport model; the upper and lower bounds account for the uncertainty in the width of the suspended graphene under high bias (see Supplementary Section 8 for details).

# Supplementary Information

## Table of contents

## Movie clips

Move clip S1 - 3: visible light emission from suspended ME graphene by microscope CCD camera.

S1: few-layer ($V_{SD}$ = 2.4V –> 2.9V -> 2.4V).

S2: multi-layer (increase pulsed $V_{SD}$ = 7.5V –> 8V).

S3: monolayer ($V_{SD}$ = 2.2V –> 2.58V -> 2.2V).

Movie clip S4: visible light emission from suspended ME few-layer graphene by the digital camera (iPhone 4).

Movie clip S5: visible light emission from suspended CVD graphene under vacuum by microscope CCD camera ($V_{SD}$ = 2.2V –> 2.7V -> 2.2V).

Movie clip S6: visible light emission from suspended ME monolayer graphene with pulsed voltage bias under vacuum by microscope CCD camera (Pulsed voltage bias).

# S1. Fabrication process and characterization of suspended graphene devices

## A. Suspended mechanically exfoliated (ME) graphene

The suspended mechanically exfoliated (ME) graphene structures are realized with nano-fabrication processes utilizing a PMMA mediated micro contact transfer method as shown in Supplementary Fig. 1.

(1) Mechanically exfoliated graphene was prepared on a $SiO_2$/Si substrate.

(2) A PMMA (Polymethyl methacrylate, 950K C4) was spin coated on graphene at 4500 rpm followed by baking process at 180°C for 5 minutes.

(3-4) To make a patterned graphene array, PMMA on unwanted areas of graphene was exposed by e-beam lithography, and the remaining PMMA after development acted as an etch mask during $O_2$ plasma etching.

(5) Patterned graphene array was prepared after removing PMMA with acetone.

(6) PMMA was spin coated again on the patterned graphene ribbons using the same recipe as in step (2).

(7) PMMA membrane with graphene ribbons was separated from $SiO_2$/Si substrate in 10% (wt) potassium hydroxide water solution (KOH).

(8) The separated PMMA membrane with attached graphene was rinsed with DI-water to remove the KOH residue from the graphene surface and dried at room temperature in Nitrogen atmosphere.

(9) The position of the PMMA membrane with patterned graphene arrays was manipulated on prepared trench substrate (depth: 300 ~ 1100 nm) using home-made micro-position aligner.

(10) Using micro contact transfer method, each side of the graphene ribbons were attached to the Au electrodes of the prepared trench.

(11) The PMMA layer was removed by an acetone wash followed by an IPA (Isopropanol) rinse. The suspended ME graphene devices are completed after a critical point drying process (see Supplementary Figs. 2 a-f).

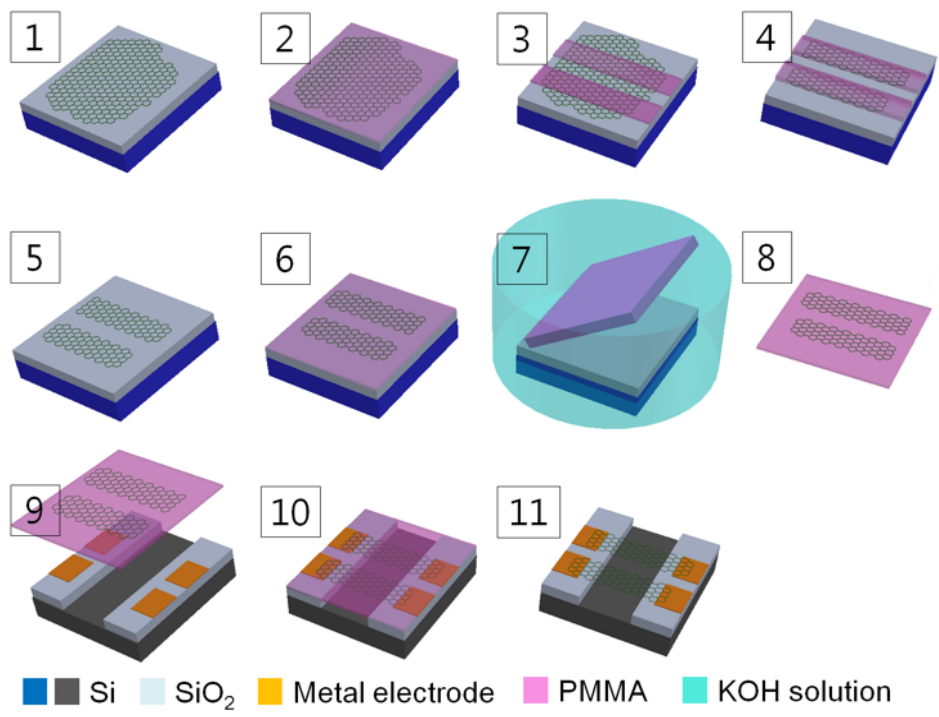

**Supplementary Fig. 1 | Schematic of fabrication process of suspended ME graphene.**

## B. Large-scale suspended CVD graphene

The large-scale suspended CVD graphene structures are realized using wet-etching method. Here, we use the large-scale LPCVD monolayer graphene initially grown on Cu foil and then transferred onto a $SiO_2$/Si substrate. In addition, we also use the large-scale plasma-assisted CVD few-layer graphene[1], which enables direct growth of few-layer graphene on an arbitrary substrate without transfer and stacking process.

(1) A PMMA (950K C4) was spin coated on CVD graphene with at 4500 rpm followed by baking process at 180°C for 5 minutes.

(2) To make a patterned CVD graphene array, PMMA on unwanted areas of graphene was exposed by e-beam lithography, and the remaining PMMA after development was acted as an etch mask during $O_2$ plasma etching.

(3) Patterned graphene array was prepared after removing PMMA with acetone.

(4) Electrodes were patterned by e-beam lithography.

(5) Metal deposition (Cr/Au=20/80 nm) and lift-off.

(6) $SiO_2$ was removed using buffered oxide etchants (BOE) or HF and rinsed in D.I. water.

(7) The large-scale suspended CVD graphene devices are completed after critical point drying process (see Supplementary Figs. 2 g-h).

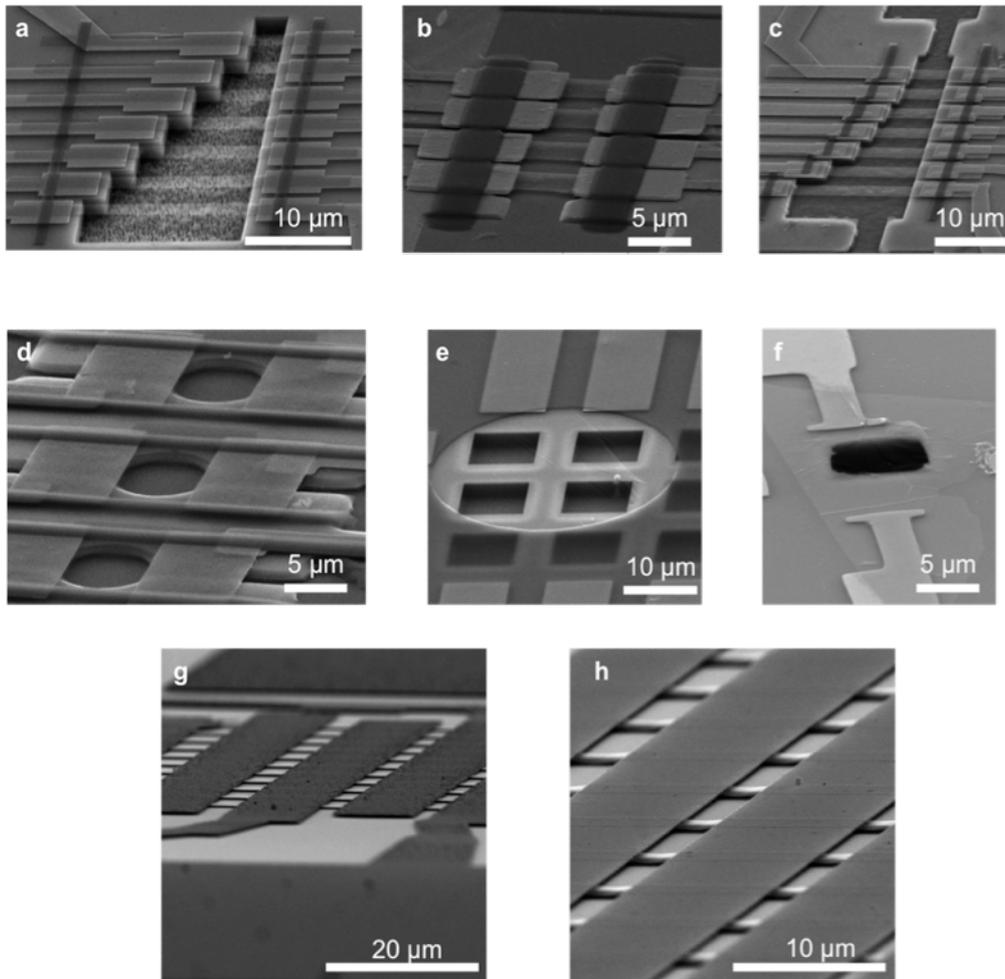

**Supplementary Fig. 2 | Scanning electron microscopy images of suspended ME graphene devices (a-f) and large-scale suspended CVD graphene devices (g-h).**

## C. Characterization of suspended graphene devices.

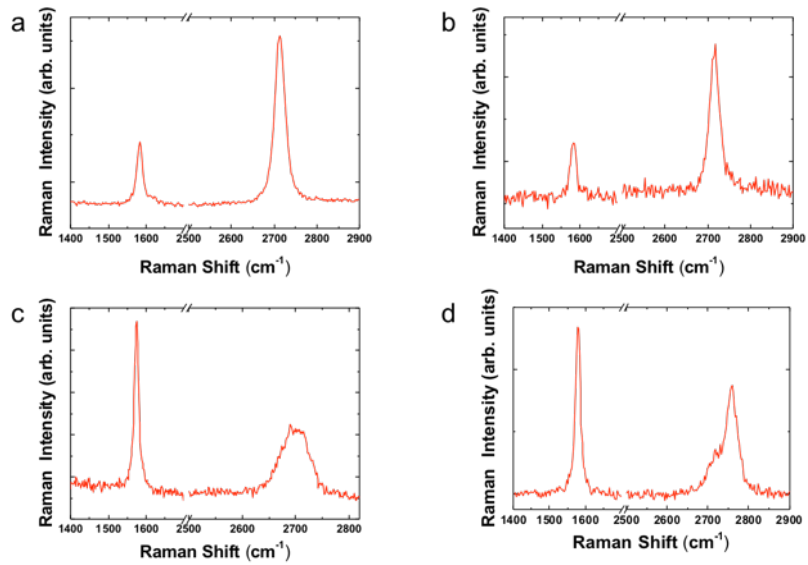

**Supplementary Fig. 3 | Raman spectroscopy of suspended ME graphene devices. a** and **b,** Two different monolayer graphene devices. **c,** Tri-layer graphene. **d,** Multi-layer graphene.

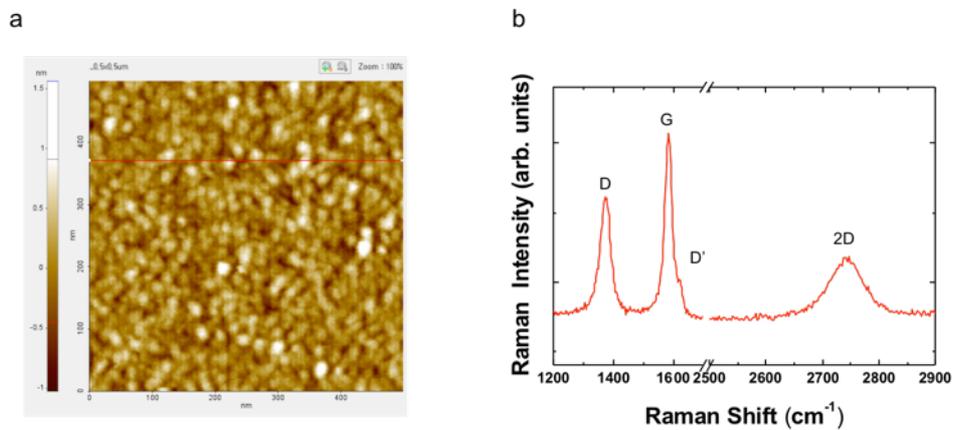

**Supplementary Fig. 4 | Characterization of plasma-assisted CVD direct grown few-layer graphene. a,** AFM image of unsuspended CVD few-layer graphene on $SiO_2$/Si substrate, we estimate the grain size to be ~ 20 nm and the average thickness ~ 1nm. **b,** Raman spectroscopy of suspended CVD few-layer graphene devices, D and D' peaks are attribute to the small grain size.

## S2. Current-induced annealing for high quality contact

As-fabricated suspended ME graphene devices usually have a high total resistance from a few tens to hundreds of kilo ohms, which is largely due to the transfer process of patterned graphene to a pre-fabricated trench and electrodes accompanied by PMMA residue and contaminants on the channel and at the interface between graphene/metal electrodes. These impurities become the dominant scattering sources for mobile carriers and increase the contact resistance of suspended ME devices[2]. To achieve a clean graphene channel as well as stable and reliable high quality electrical contacts, we swept the source-drain bias voltage with slow increases in the maximum bias voltage at each step. While sweeping the bias voltage, we observe a significant decrease in the total resistance of suspended ME graphene devices as shown in Supplementary Fig. 5. It is attributed to the slow current-induced annealing effect, which removes the impurities at the graphene/metal electrode interfaces as well as graphene channel, resulting in an improvement in the electrical contact quality. After the annealing process, the contact resistivity (1.2 ± 0.2 kΩ·μm) for the suspended monolayer graphene was obtained by the transfer length method[3] as shown in Supplementary Fig. 6.

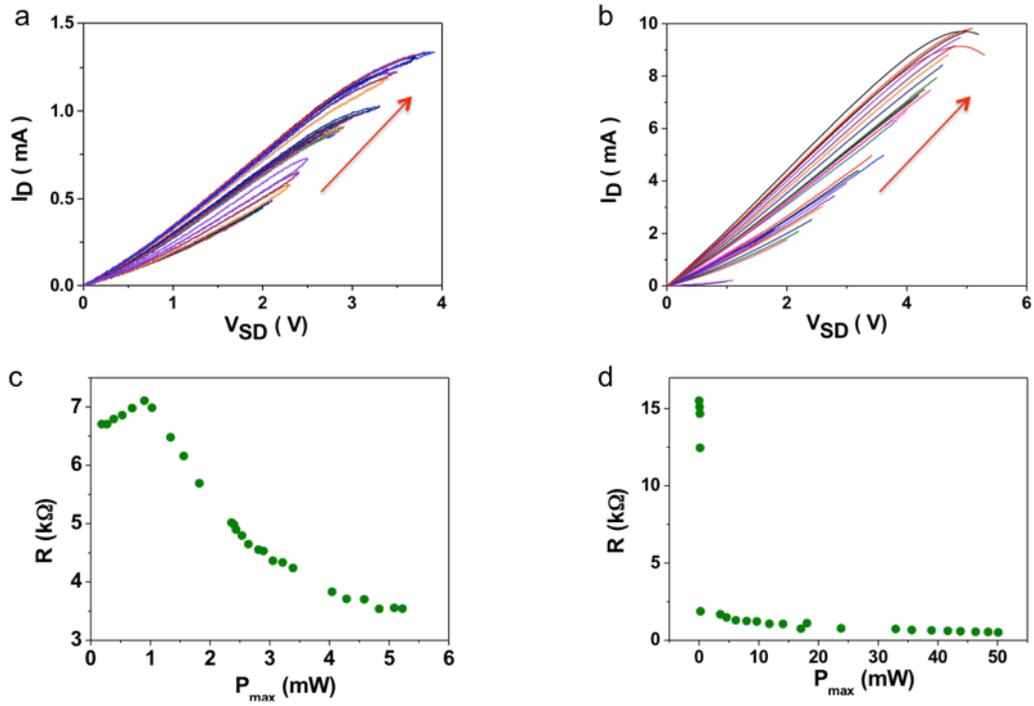

**Supplementary Fig. 5 | Total resistance reduction by repetitive sweeping of the source-drain bias voltage**. *I-V* curve of suspended ME (a) tri-layer graphene ($L$ =6.5 um, $W$ = 3 um) and (b) multi-layer graphene ($L$ =9 um, $W$ = 3 um). The total resistance of suspended graphene obtained at $V_{SD}$ = 0.1 V as a function of maximum power $P_{max}$ by repetitive sweeping of $V_{SD}$ (where the maximum value of $V_{SD}$, $V_{max}$, increases in each sweeping cycle) is lowered and approaches (c) ~ 3.5 kΩ (tri-layer) and (d) ~ 0.5 kΩ (multi-layer). Reduction in the total resistance of suspended graphene is attributed to the current-induced annealing of graphene/metal interface as well as graphene channel, which removes fabrication-process-induced contaminants such as resist residue, resulting in an improvement in the quality of contacts and graphene channel. Arrows in panels (a) and (b) indicate the time evolution of the *I-V* curves.

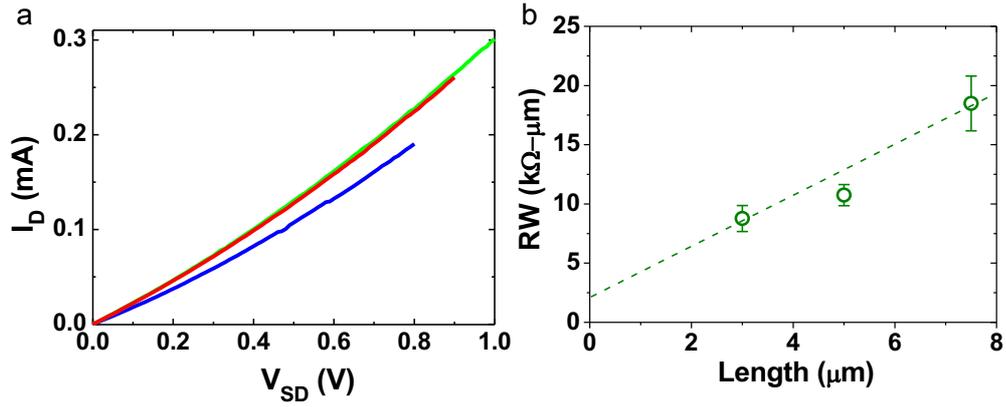

**Supplementary Fig. 6 | Contact resistance of suspended ME monolayer graphene. a**, *I-V* curves of a suspended ME monolayer graphene ($L = 3$ μm, $W = 2$ μm) in different sweeping cycles. The resistance is lowered if the maximum value of $V_{SD}$ is higher than ~ 0.8 V, as seen in the evolution from the blue curve to the red one. **b**, After current-induced annealing process, the contact resistivity ($R_C W$) of ~ 1.2 kΩ·μm of suspended ME monolayer graphene devices was estimated from the transfer length method: the *y*-axis intersecting value of the extrapolated line for the three data points corresponds to $2R_C W$. Here, the error bar was deduced from the uncertainty of graphene width mainly due to the fabrication process ($W \approx 2 \pm 0.25$ μm).

## S3. Stable visible-light emission from suspended graphene

We observe stable and reproducible bright visible-light emission from suspended graphene devices under modest electric field as shown in movie clips of S1 - S6 in vacuum environments. Especially, as shown in movie clips S3 and S6, suspended ME monolayer graphene exhibits the repeated visible light emission and fast pulsed light emission without degradation and failure. Supplementary Fig. 7 exhibits the reliability, durability and reproducibility of light emission from suspended monolayer graphene under modest electric field. Furthermore, the brightest light emission position of suspended graphene is consistently at the centre of graphene as shown in Supplementary Fig. 8. The centre is the location of maximum temperature of suspended graphene where the heat dissipation is minimal.

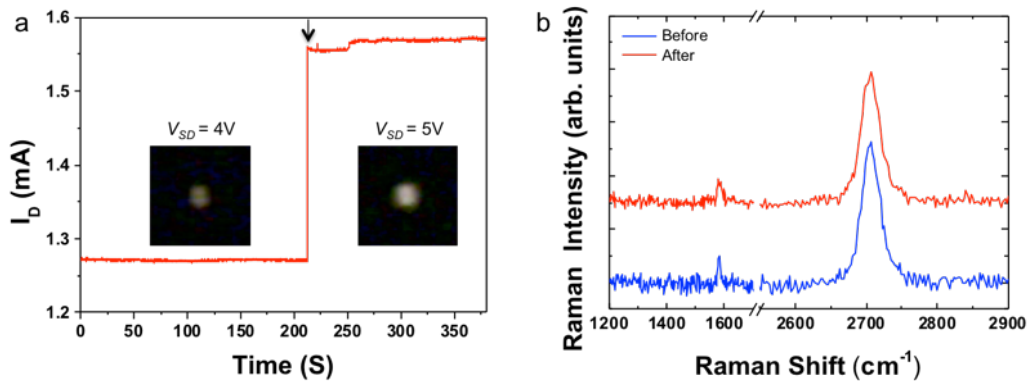

**Supplementary Fig. 7 | a**, Current as a function of time during the bright visible-light emission from suspended ME monolayer under constant source-drain bias ($V_{SD}$ = 4 V and 5 V). Arrow indicates the increase in applied bias voltage ($V_{SD}$ = 5 V). Inset, optical images of bright visible light emission at the centre of suspended graphene under the two bias voltages. **b**, Raman spectra of another suspended monolayer graphene before and after bright visible light emission. Note that both spectra do not have the D peak induced by defects even after bright visible light emission (spectrum after light emission was vertically shifted for clarity).

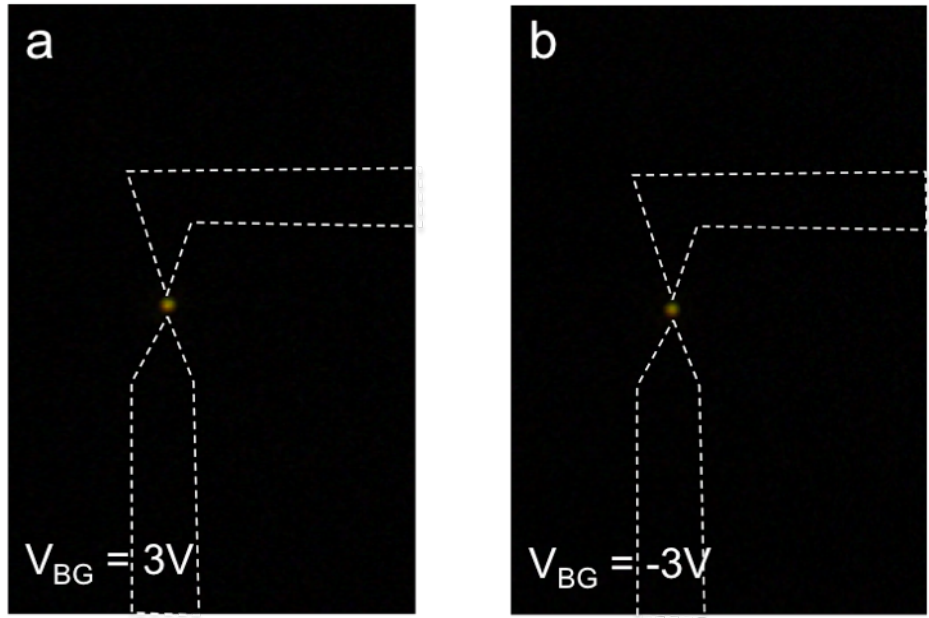

**Supplementary Fig. 8 | Light emission spot with different gates.** Optical image of bright visible light emission from electrically biased suspended CVD monolayer graphene under vacuum probe station. Here, dashed lines indicate the outlines of electrodes. We applied bias voltage ($V_{SD}$ = 2.8V, $I_D$ = 465 μA) across the suspended graphene ($L$ = 7 μm, $W$ = 7 μm) with different gate voltages. We observe that the brightest spot consistently remains at the centre of suspended graphene regardless of the gate voltage.

## S4. Optical measurements and spectrum analysis

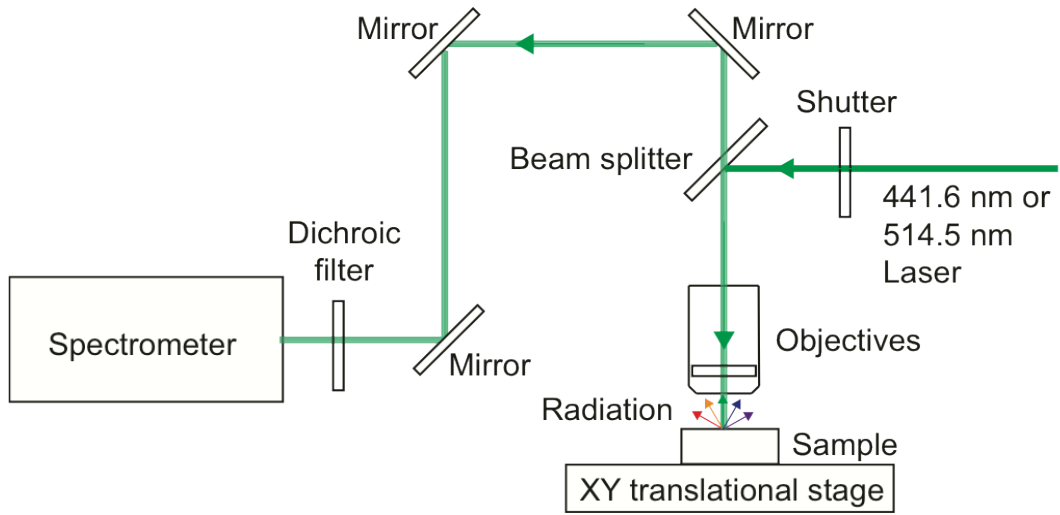

**Supplementary Fig. 9 | Schematic illustration of the optical measurement set-up for the observation of bright visible-light emission and the acquisition of Raman spectra from electrically biased suspended graphene.**

The Raman spectra were measured using the 514.5 nm line of an Ar ion laser or the 441.6 nm line of a He-Cd laser with a power of 500 μW. We used a 50× objective lens (NA 0.42 and WD 20.3 mm) to focus the laser beam onto the sample, which was housed in a vacuum of $<10^{-4}$ Torr at room temperature. A Jobin-Yvon Triax 320 spectrometer (1200 groove/mm) and a charge-coupled device (CCD) array (Andor iDus DU420A BR-DD) were used to record the spectra. The bright visible-light emission spectra were measured using the same system. At each bias voltage, the Raman and light-emission spectra were measured sequentially, using a motorised flipper mount for a dichroic filter and an optical beam shutter (Thorlabs SH05). The throughput of the optical system was carefully calibrated using a calibrated black-body source (1255 K, OMEGA BB-4A) and a tungsten filament (3000 K, which were calibrated against the International System of Units the Korea Research Institute of Standards and Science).

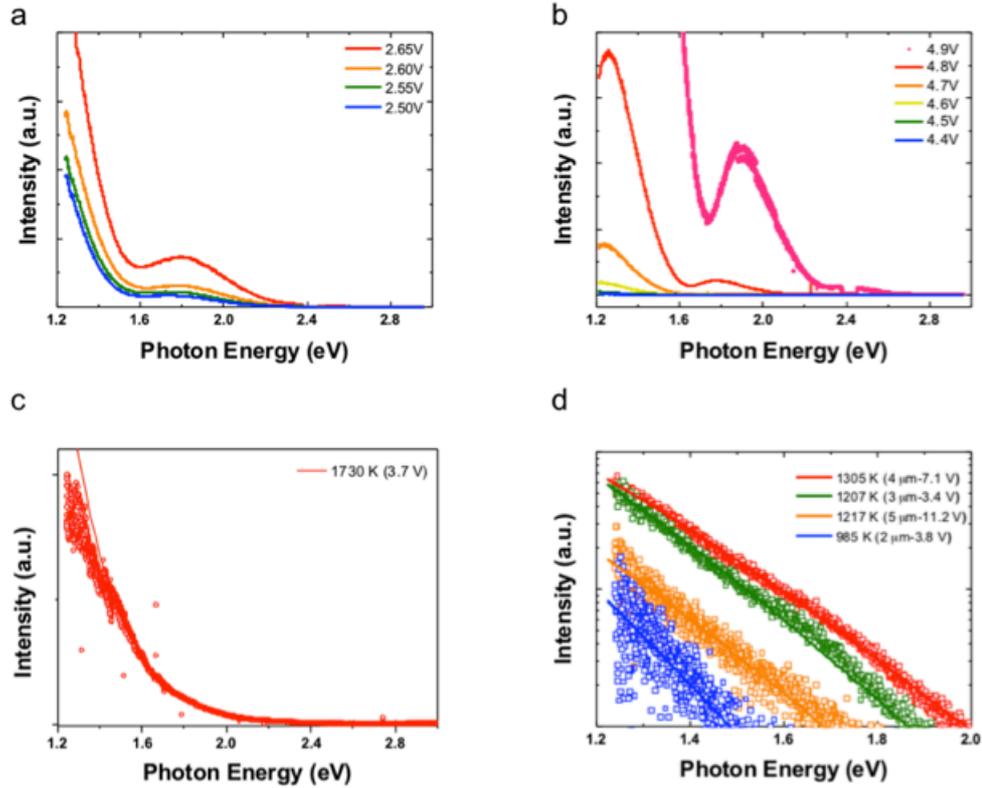

**Supplementary Fig. 10 | Spectra of visible-light emission from electrically biased suspended graphene with various trench depths. a** and **b,** Spectra of devices with trench depths of $D = 800 \sim 1000$ nm consisting of suspended (**a**) tri-layer ($L = 12$ μm, $W = 3$ μm) and (**b**) multi-layer ($L = 9$ μm, $W = 3$ μm) graphene for various bias voltages, exhibiting multiple strong light-emission peak structures because of the strong interference effect. **c** and **d**, Spectra of devices with shallow trenches consisting of suspended (**c**) CVD monolayer graphene with a trench depth of ~ 80 nm (linear scale) and (**d**) CVD few-layer graphene with a trench depth of ~ 300 nm (log scale), exhibiting grey-body-like features (solid curve).

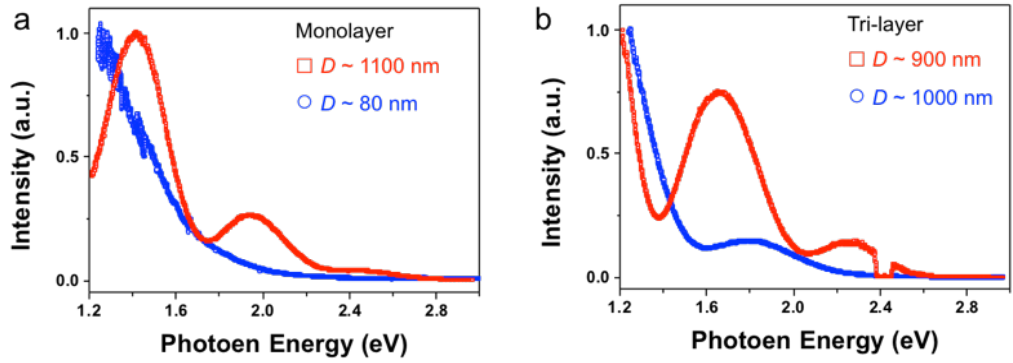

**Supplementary Fig. 11 | Comparison of measured radiation spectra of suspended graphene with different trench depths. a,** Normalized radiation spectra of suspended monolayer graphene with different trench depths (red square $D \sim 1100$ nm from Fig. 2a, blue circle $D \sim 80$ nm from Supplementary Fig. 10c). **b**, Normalized radiation spectra from suspended tri-layer graphene with different trench depths (red square $D \sim 900$ nm from Fig. 2b, blue circle $D \sim 1000$ nm from Supplementary Fig. 10a). The abrupt dip in the radiation spectrum of tri-layer graphene for $D \sim 900$ nm is due to the dichroic filter. These results are consistent with theoretical simulation results in Fig. 3c.

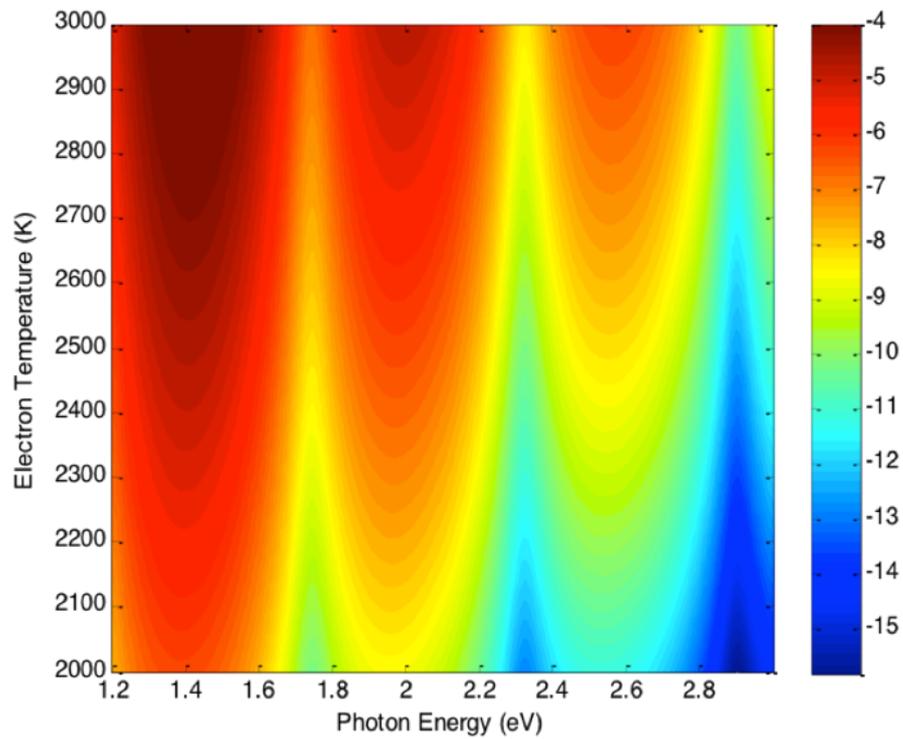

**Supplementary Fig. 12 | Simulated spectra based on the interference effect of thermal visible radiation from electrically biased suspended ME monolayer graphene.** Spectrum modulation achieved by varying the electron temperature of suspended monolayer graphene at a constant trench depth (d = 1070 nm, corresponding to Fig. 2a).

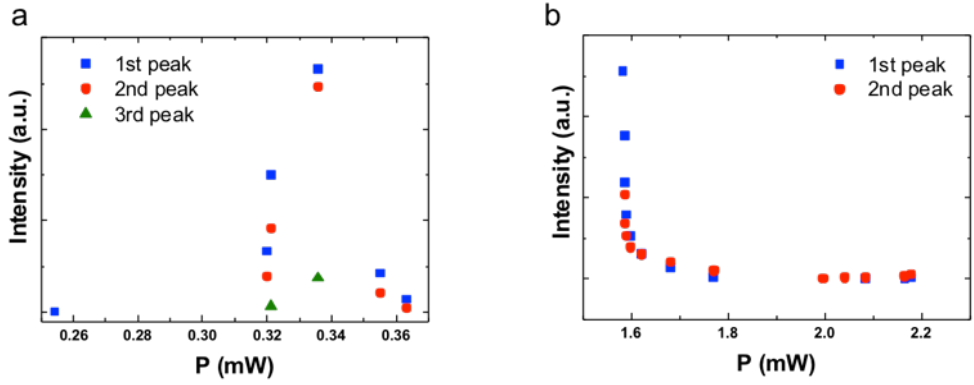

**Supplementary Fig. 13 | Radiation intensity as function of applied electrical power.** Intensity of light emission from electrically biased suspended ME (**a**) monolayer (**b**) tri-layer graphene, corresponding to Figs. 2c and 2d, respectively, versus applied electrical power. The results clearly demonstrate that the intensity of light emission from suspended graphene is strongly correlated with the applied electric field rather than the applied electric power. The observed decrease in the electrical power with increasing bias voltage is attributed to thermal annealing or etching effect on the graphene channel at high temperatures.

## S5. Beyond black-body radiation formula for visible light emission from suspended graphene: Dependence on the number of graphene layers

When photons are in equilibrium with matter at temperature $T$, the intensity spectrum of light obeys black-body radiation formula

$$I(\omega) \propto \omega^3/[\exp(\hbar\omega/k_{\text{B}}T) - 1]$$

being independent of electronic structure of the material. In describing thermal emission spectra of suspended graphene layers, however, we cannot assume that photons are in equilibrium with the system because photons escape the system immediately as soon as they are emitted by electron-hole recombination, i.e., they are not absorbed again and generate electron-hole pairs. For this reason, use of black-body radiation formula to describe electroluminescence from graphene samples is not well grounded. However, as we discuss below, results based on black-body radiation formula are quite a good approximation except in the case of low-energy radiation (the criterion for low-energy regime depends on the temperature), which is why all the previous studies have used this formula to interpret electroluminescence from graphene samples.

In order to address the number-of-layers dependence of the emission spectra, one needs a theory beyond the black-body radiation formula that takes care of the actual electronic structure of the system. Using the Fermi golden rule, the intensity of spontaneously emitted light per frequency per solid angle in the direction $\hat{n}$ from a material is given by [4]

$$I(\omega) \propto \omega^2 \sum_{snm\mathbf{k}} |\langle n\mathbf{k}|\mathbf{p}\cdot\boldsymbol{\epsilon}_s|m\mathbf{k}\rangle|^2 \delta(E_f - E_i - \hbar\omega) f_{\text{FD}}(E_i)\left(1 - f_{\text{FD}}(E_f)\right)$$

where $|m\mathbf{k}\rangle$ and $|n\mathbf{k}\rangle$ are the initial and final Bloch energy eigenstates with energy eigenvalues $E_i$ and $E_f$, respectively, $\boldsymbol{\epsilon}_s$ polarization vector normal to $\hat{n}$, and $f_{\text{FD}}(E) = [\exp(E/k_{\text{B}}T) + 1]^{-1}$, the Fermi-Dirac occupation at temperature $T$. This formula describes light emission arising from direct radiative transitions, emitting photons to vacuum in $\hat{n}$ direction (which is the surface-normal direction in our experiment).

To obtain the emission intensity, we calculated the electronic band structure of graphene with various geometries within a simple tight binding model. In this scheme, the energy band is obtained by diagonalising the $\mathbf{k}$-dependent Hamiltonian $H(\mathbf{k}) = \exp(-i\mathbf{k}\cdot\mathbf{r})\,H\,\exp(i\mathbf{k}\cdot\mathbf{r})$, where $H$ denotes the Hamiltonian describing nearest-neighbor intra-layer and vertical interlayer hopping between local $2p_z$ atomic orbitals with hopping integrals -2.8 eV and 0.4 eV, respectively.

The momentum matrix element $\mathbf{p}_{nm}(\mathbf{k}) = \langle n\mathbf{k}|\mathbf{p}|m\mathbf{k}\rangle$ is given by[5]

$$\mathbf{p}_{nm}(\mathbf{k}) = \frac{m_e}{i\hbar}\langle n\mathbf{k}|[\mathbf{r}, H]|m\mathbf{k}\rangle = \frac{m_e}{i\hbar}\langle u; n\mathbf{k}|[\mathbf{r}, H(\mathbf{k})]|u; m\mathbf{k}\rangle$$

$$= \frac{m_e}{\hbar}\left\langle u; n\mathbf{k}\left|\frac{\partial}{\partial \mathbf{k}}H(\mathbf{k})\right|u; m\mathbf{k}\right\rangle,$$

where $m_e$ is the mass of an electron and $|u; n\mathbf{k}\rangle$ the periodic part of the Bloch state $|n\mathbf{k}\rangle$ (i.e. $|n\mathbf{k}\rangle = \exp(i\mathbf{k}\cdot\mathbf{r})|u; n\mathbf{k}\rangle$).

Supplementary Fig. 14 shows the simulated emission spectra of graphene with different numbers of layers and stacking sequences at 2500 K. A black-body radiation spectrum (with an overall scaling factor to match the other spectra at high-energies) is shown for comparison. For the low-energy part, the calculated emission spectra from graphene deviate from the black-body radiation formula. Bi-layer and tri-layer graphene exhibit characteristic peaks near 0.4 ~ 0.6 eV corresponding to the direct transition at K point. In addition, for monolayer graphene,

$$I(\omega) \propto \frac{\omega^3}{[\exp(\hbar\omega/2k_\mathrm{B}T) + 1]^2} \underset{\hbar\omega \ll k_\mathrm{B}T}{\sim} \omega^3$$

in the low-energy regime, because (i) the energy-versus-momentum relation is linear and thus the magnitude of the momentum matrix element is constant and (ii) the density of states is linear in energy. This result should be contrasted with the ordinary black-body radiation formula

$$I(\omega) \propto \frac{\omega^3}{\exp(\hbar\omega/k_B T)-1} \underset{\hbar\omega \ll k_B T}{\sim} \omega^2,$$

which shows different low-energy behaviors from the correct theory for graphene.

However, for energies higher than 1.2 eV (at $T$=2500 K), all the spectra become similar and can be fitted well by the black-body radiation formula because (i) the interlayer coupling modulates the electronic structure mostly in the low-energy regime (roughly within the interlayer hopping integral), and (ii) the emission spectrum in high-energy regime is essentially determined by the Fermi-Dirac occupation factors $f_{\text{FD}}(E/2)(1 - f_{\text{FD}}(-E/2))$, which can be approximated by the Boltzmann factor $\exp(-E/k_B T)$. In the experiment we measure emission spectra in the photon energy from 1.2 eV to 3.0 eV; hence, the measured spectra are indistinguishable from the black-body radiation spectrum, irrespective of the number of layers and stacking sequence.

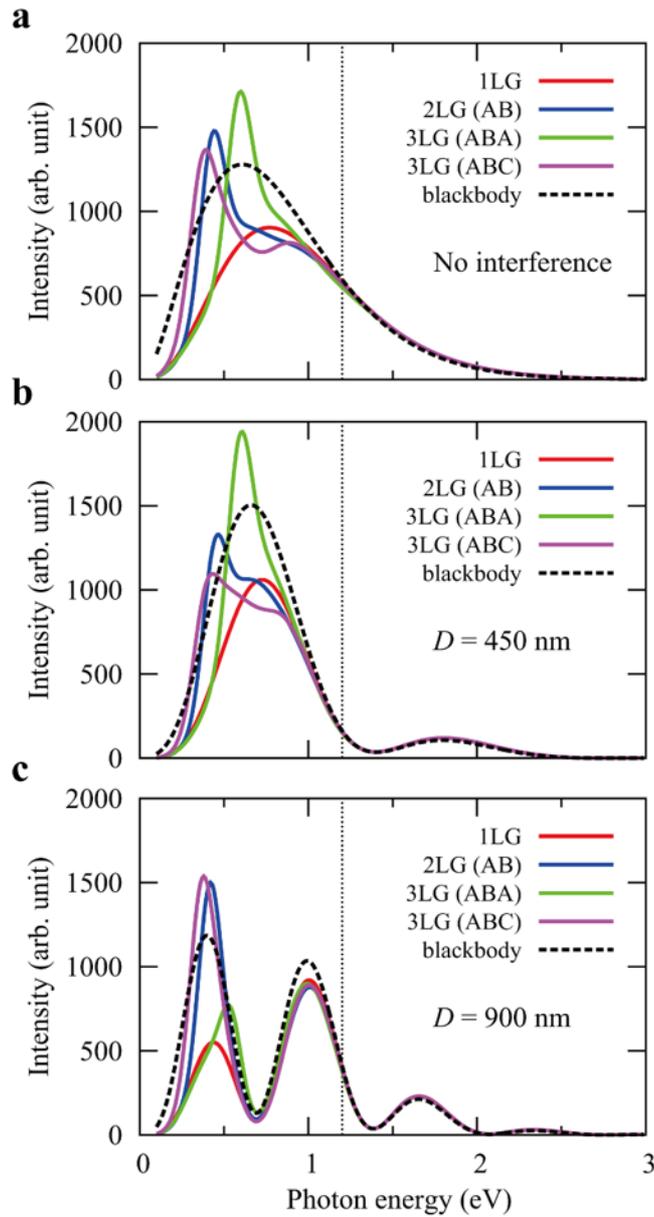

**Supplementary Fig. 14 | Calculated direct emission intensity at $T = 2500$K for monolayer graphene (1LG), bi-layer graphene (2LG) with AB stacking and tri-layer graphene (3LG) with ABA and ABC stacking. a,** Intrinsic direct emission spectrum, **b,** spectrum with interference effect taken into account for trench depth $D = 450$ nm and **c,** similar quantities as in (b) with $D = 900$ nm. The black-body radiation curve $I(E) = I_0 E^3/[\exp(E/k_\mathrm{B}T) - 1]$ is also shown for comparison in each panel. The intensities are divided by the number of layers for normalization

# S6. Effects of absorption and reflection by graphene layers on the emission spectra

It is known that monolayer graphene absorbs as much as 2.3% of light[6]. In this section we study the effects of absorption and reflection by graphene, both monolayer and tri-layer, on the emission spectra. We used the method described in D. Yoon *et al.*[7] to incorporate the effect of absorption and reflection by graphene layers into our calculation. The thickness of graphene is assumed to be 0.335 nm per layer and the refractive index of graphene is assumed to be $n(hv) = 2.6 + 2.66i/(hv/\text{eV})$, whose real part is taken from the corresponding value of graphite and the imaginary part is chosen to ensure 2.3% light absorption per each graphene layer.

Absorption and reflection by graphene layers themselves, in the case of tri-layer graphene could make as large as ~4% difference in the calculated emission intensity (Supplementary Fig. 15d). However, the differences at intensity maxima (see Supplementary Fig. 15c) are rather small (Supplementary Fig. 15d) and hence the absolute difference in the emission spectra arising from the absorbance and reflectance by graphene layers is small (see the green dash-dotted curves in Supplementary Figs. 15a and 15c).

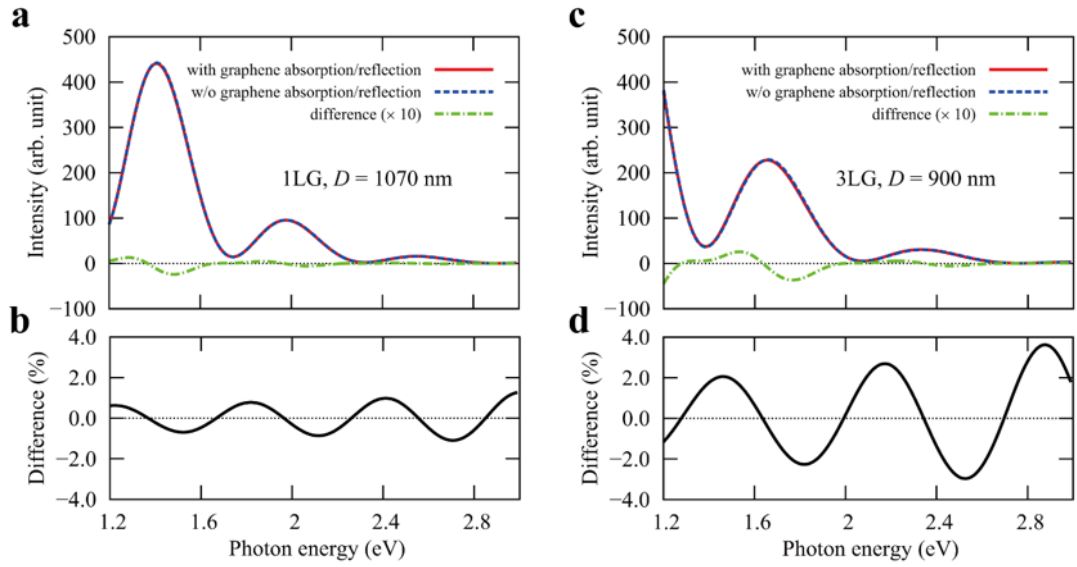

**Supplementary Fig. 15 | a,** The simulated emission spectrum of monolayer graphene with trench depth $D$ = 1070 nm. The red solid curve (blue dotted curve) shows the spectrum considering (neglecting) absorption and reflection by graphene. The green dash-dotted curve shows the (magnified) difference between the two spectra. **b,** The relative increase in the intensity made by incorporating the effect of absorption and reflection by graphene. **c** and **d**, Similar quantities as in (a) and (b) for tri-layer ABA graphene with trench depth $D$ = 900 nm.

## S7. Negative differential conductance (NDC) and hysteresis behavior in electrically biased suspended ME graphene

Interestingly, an NDC behavior is exhibited by electrically biased suspended ME few-/multi-layer graphenes as shown in Supplementary Figs. 16 and 17. The critical electric field (0.40 ~ 0.43 V/µm) of suspended few-layer graphene for activation of intrinsic graphene OPs is not changed during multiple sweeps and current decreasing hysteresis, which could be related to the thermal annealing effect or a narrowing of channel width at high temperature, as shown in Supplementary Figs. 16 and 17, which indicates that the hot OPs of graphene are dominantly populated by the applied high electric field. Furthermore, we can observe that the initiation of visible light emission from electrically biased suspended graphene occurs near the critical electric field and across region of zero differential conductance. It may be attributed to the accelerated charge carriers in electric fields obtaining enough energy for emission of intrinsic graphene OPs[8,9], causing strong electron-OP scattering and NDC behavior in suspended graphene, which is likely dependent upon the applied electric field rather than the current flow or electrical power dissipation.

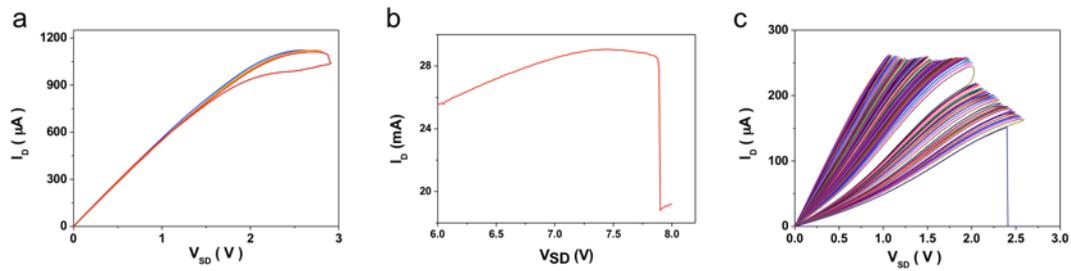

**Supplementary Fig. 16 | Electrical transport data corresponding to optical images of bright visible light emission from suspended ME graphene in Fig. 1.** Electrical transport data of suspended ME (**a**) few-layer (**b**) multi-layer (**c**) monolayer graphene corresponding Figs. 1d-f. Initiation of visible light emission from suspended ME graphene are observed at bias voltage corresponding to zero differential conductance.

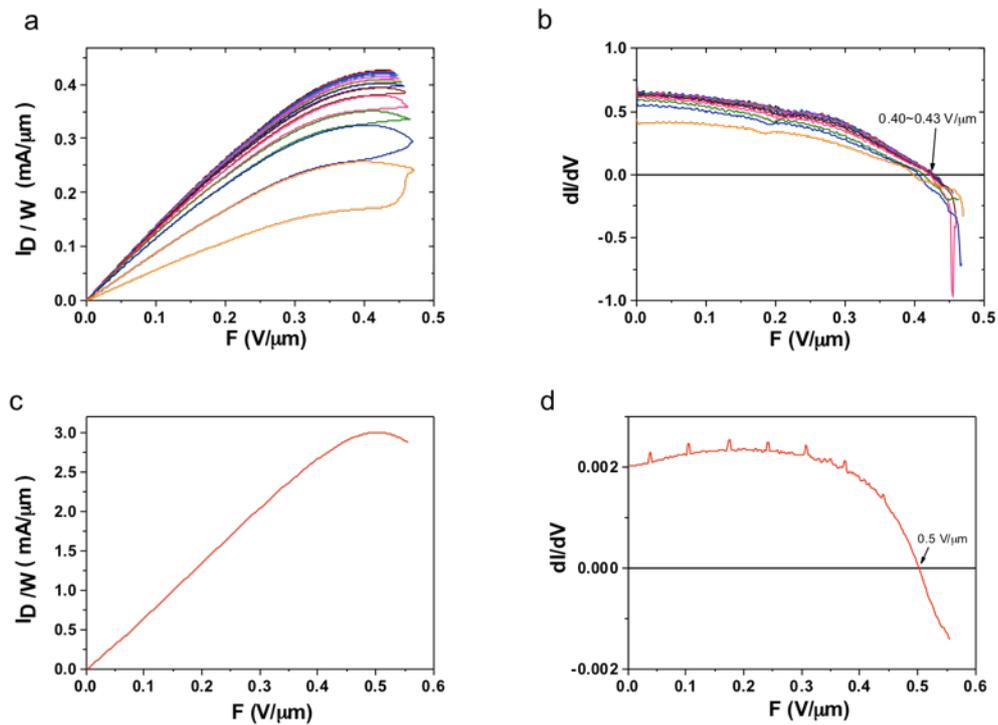

**Supplementary Fig. 17 | Current saturation and negative differential conductance (NDC) in electrically biased suspended graphene**. **a and c,** Electrical transport ($I_D/W$ vs. $F$ curve) and differential conductance of suspended ME (**a** and **b**) few-layer graphene ($L$ = 6.5 μm, $W$ = 3 μm) and (**c** and **d**) multi-layer graphene ($L$ = 9 μm, $W$ = 3 μm). Current saturation and negative differential conductance occur at electric fields higher than a certain critical electric field (0.40 ~ 0.43 V/μm for few-layer graphene and 0.5 V/μm for multi-layer graphene); this behaviour is attributed to the electric-field-induced accumulation of hot electrons and OPs in the suspended graphene.

# S8. Intrinsic electro-thermal transport model in suspended graphene

The current density $J (= I_D / W)$ in a graphene is expressed by the continuity equation,

$$J = e(n_{ex} + n_{px})v_{dx} \tag{1}$$

, where $e$ is the elementary charge, $n_e$ ($n_p$) is the electron (hole) density, $v_d$ is the drift velocity and the subscript $x$ is the location along the graphene channel ($x = 0$ at the middle of the channel). Thermally excited carrier densities ($n_{th}$) for the mono- and tri-layer graphenes are different from each other due to their different band structures. In a case of monolayer graphene with a linear band structure,

$$n_{th} = \frac{\pi}{6}\left(\frac{k_B T_x}{\hbar v_F}\right)^2 \tag{2}$$

, where $k_B$ is the Boltzmann constant, $\hbar$ Planck constant and $v_F$ ($\sim 10^6$ m/s) the Fermi velocity. For a tri-layer graphene with a parabolic band structure,

$$n_{th} = \frac{2m^*}{\pi \hbar^2} k_B T_x \ln(2) \tag{3}$$

, where $m^*$ ($=0.082 m_e$) is the effective mass of electron in a tri-layer graphene, $m_e$ the mass of electron[10]. Supplementary Fig. 18 shows the total thermally excited carrier density as a function of temperature for the mono- and tri-layer graphenes.

The drift velocity is expressed by

$$v_{dx} = \frac{\mu_x F_x}{[1+(\mu_x F_x/v_{sat})^\eta]^{1/\eta}} \tag{4}$$

, where $\mu_x$ is the temperature-dependent mobility and $F_x = -dV_x/dx$ is the electric field along the graphene, $v_{sat}$ is a saturation velocity related to the OP scattering and $V_x$ is the local potential along the graphene. Here, we set $\eta \equiv 2$[8,9,11]. With above electrical expressions, we add a heat diffusion equation to obtain the temperature profile along the graphene:

$$\frac{d^2 T_x}{dx^2} + \frac{1}{\kappa_x W t}\frac{dP_x}{dx} - \frac{2g_x}{\kappa_x t}(T_x - T_0) = 0 \tag{5}$$

, where $P_x$ [$= I(V_x - IR_c)$] is the locally dissipated power within the suspended channel, $\kappa_x$ the temperature-dependent thermal conductivity of the suspended graphene, $L$ the length, $W$ the width, $t$ the thickness of the graphene, and $T_0$ the base temperature. Here, $g_x$ is the thermal conductance per unit area between graphene and environment, e.g., $g_x = 0$ for $-L/2 < x < L/2$ (a suspended region). The contact resistivity, $\rho_c = 2 \times 10^{-5}$ $\Omega \text{cm}^2$ was considered for further calculations.

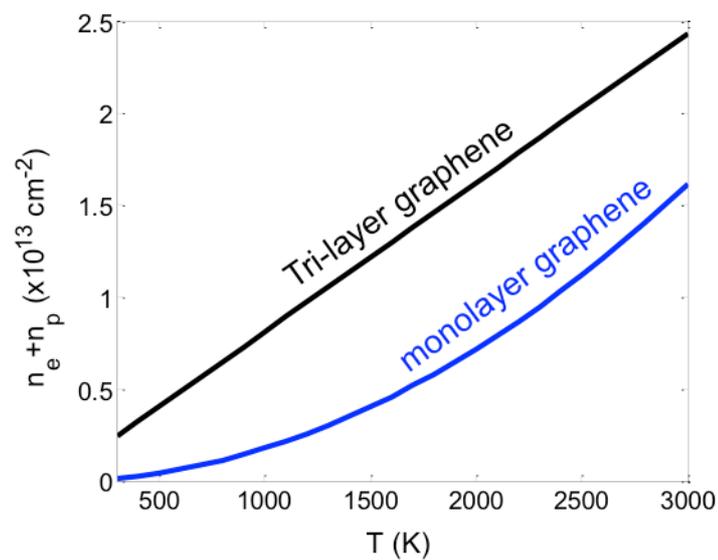

**Supplementary Fig. 18 | Total thermally excited carrier density as a function of temperature for the monolayer and tri-layer graphenes.**

## A. Tri-layer graphene

For reliable simulations with many parameters in the calculation process, one needs lattice temperature (or $T_{ap}$) information obtained from the experiments as well as $I_D$-$V_{SD}$ curves. To get the temperature information from the tri-layer graphene device, a frequency of the G peak in the Raman spectroscopy with various biases was measured as shown in Supplementary Fig. 19a. The frequency of the G peak shows a downshift with increasing $V_{SD}$ (denoted near corresponding data in the figure). The downshift of the G peak is related to an enhancement of anharmonic phonon-phonon coupling with increasing temperature and provides temperature of lower energy (secondary) phonons, $T_{sp}$[10,12]. Based on Ref. S10, we estimated $T_{sp}$ as a function of $V_{SD}$ (scattered grey points) in Supplementary Fig. 19d. We also plot $T_e$ as a function of $V_{SD}$ (opened scattered diamonds) in Supplementary Fig. 19d based on the temperatures estimated from the thermal emission in Fig. 2b. We could not distinguish the Raman spectrum for $V_{SD}$ > 3.35 V because the light emission intensity at photon energy ~ 2.4 eV corresponding to the Raman laser wavelength was larger than the Raman signal for the bias conditions. Now, we simulate the corresponding $I_D$-$V_{SD}$ curves showing current saturation behavior as shown in the inset of Supplementary Fig. 19d based on the electro-thermal model, where green and red scattered points (solid curves) are data (calculation results) for two thermal emission conditions of $V_{SD}$ = 3.35 V and 3.65 V (indicated by green and red arrows), respectively. Here, we consider that the different slope between the two curves could be mainly due to an edge burning effect at high temperatures during the measurement, resulting in narrower widths. We also numerically get the temperature and thermal-conductivity profiles as shown in Supplementary Figs. 19b (only for $V_{SD}$ = 3.35 V and 3.65 V) and 19c, respectively. Supplementary Table 1 shows used parameters including width ($W$) for various max-$V_{SD}$. The first and second values of $W$ for each $V_{SD}$ provide upper lower and upper bounds of temperatures ($T_{op}$ and $T_{ap}$), which are plotted in Supplementary Fig. 19d and Fig. 4d. To determine a possible minimum value of $W$, we assume that the suspended tri-layer graphene is broken when $T_{ap}$ reaches to a burning $T_b$ (~ 2300 K) of suspended tri-layer graphene. We found that $T_{ap}$ reaches to the $T_b$ at $V_{SD}$ = 3.65 V when $W$ > 1.63 μm, thus, we consider $W$ = 1.63 μm as the minimum $W$ value at the $V_{SD}$ condition. The mobility at room temperature is also determined by the given minimum $W$ as $\mu$ = 2500 cm$^2$V$^{-1}$s$^{-1}$, which is adopted as an upper bound for the mobility through entire calculations. On the other hand, for an initial experiment stage at $V_{SD}$ = 3 V and $W$ = 3 μm, we get $\mu$ = 2220 cm$^2$V$^{-1}$s$^{-1}$, which is adopted as a lower bound through entire calculations.

On the other hand, it has been known that $T_{ap}$ and $T_{op}$ are different from each other for suspended carbon nanotubes (CNTs) under a high electric field[13]. In that case, it has been assumed that electrons and OPs are under an equilibrium state, *i.e.*, $T_e$ = $T_{op}$. In our suspended graphene cases, we also consider the nonequilibrium state between electrons (or OPs) and APs. In that case, the temperature dependent mobility is directly related to $T_e$ (or $T_{op}$), but not to $T_{ap}$. In previous works for electrically biased-suspended CNTs, a nonequilibrium OP coefficient $\alpha$ was introduced as in the relation, $T_{op}$ = $T_{ac}$ + $\alpha(T_{ap}-T_0)$, and $\alpha$ = 2.4 was found to reproduce the experimental data[6]. In the work, for the electro-thermal calculations, the Landauer model for a one-dimensional transport in a diffusive regime was applied, where temperature-dependent mean free paths of AP and OP (including applied-voltage dependence for OP) were considered. In our model for the electrically-biased suspended graphene case in a diffusive transport regime, instead of basing the model with mean free paths, we use $T_e$-dependent mobility, $\mu(T_e)$ = $\mu(T_0)(T_0/T_e)^\beta$ based on a traditional drift velocity-field relation, where $T_0$ = 300 K. In addition, the thermally excited carrier density is also expressed by $T_e$. On the other hand, thermal conductivity of the graphene is determined by $T_{ap}$, $\kappa(T_{ap})$ = $\kappa(T_0)(T_0/T_{ap})^\gamma$. For the tri-layer case, we used $\kappa(T_0)$ = 1900 Wm$^{-1}$K$^{-1}$, $\beta$ = 1.155 and $\gamma$ = 1. We find that $T_{op}$ estimated by the electro-thermal transport model matches with $T_e$ estimated by the thermal emission model with $\alpha$ = 0.3 for 3.45 V < $V_{SD}$ < 3.656 V. Importantly, we note that the calculated $T_{ap}$ by the electro-thermal model is also consistent with $T_{sp}$ estimated by the G-peak shift for $V_{SD}$ < 3.35 V.

| $V_{SD}$ (V) | $\mu$ (cm$^2$V$^{-1}$s$^{-1}$) | W (μm) | $T_{op}$ (K) | $T_{ap}$ (K) |
|---|---|---|---|---|
| 3.65 | 2220, 2500 | 1.73, 1.63 | 2425, 2866 | 1934, 2275 |
| 3.6 | 2220, 2500 | 1.78, 1.67 | 2284, 2741 | 1826, 2177 |
| 3.55 | 2220, 2500 | 1.82, 1.71 | 2240, 2650 | 1792, 2106 |
| 3.5 | 2220, 2500 | 1.89, 1.78 | 2157, 2508 | 1729, 1999 |
| 3.45 | 2220, 2500 | 2.02, 1.89 | 2017, 2412 | 1620, 1924 |
| 3.4 | 2220, 2500 | 2.17, 2.05 | 1989, 2333 | 1600, 1865 |
| 3.35 | 2220, 2500 | 2.62, 2.46 | 1902, 2212 | 1533, 1771 |
| 3.3 | 2220, 2500 | 2.8, 2.64 | 1852, 2124 | 1494, 1704 |
| 3.25 | 2220, 2500 | 2.9, 2.72 | 1744, 2049 | 1410, 1645 |
| 3 | 2220, 2500 | 3, 2.82 | 1447, 1645 | 1182, 1334 |

**Supplementary Table 1.** Parameters used for electro-thermal simulations for the tri-layer graphene.

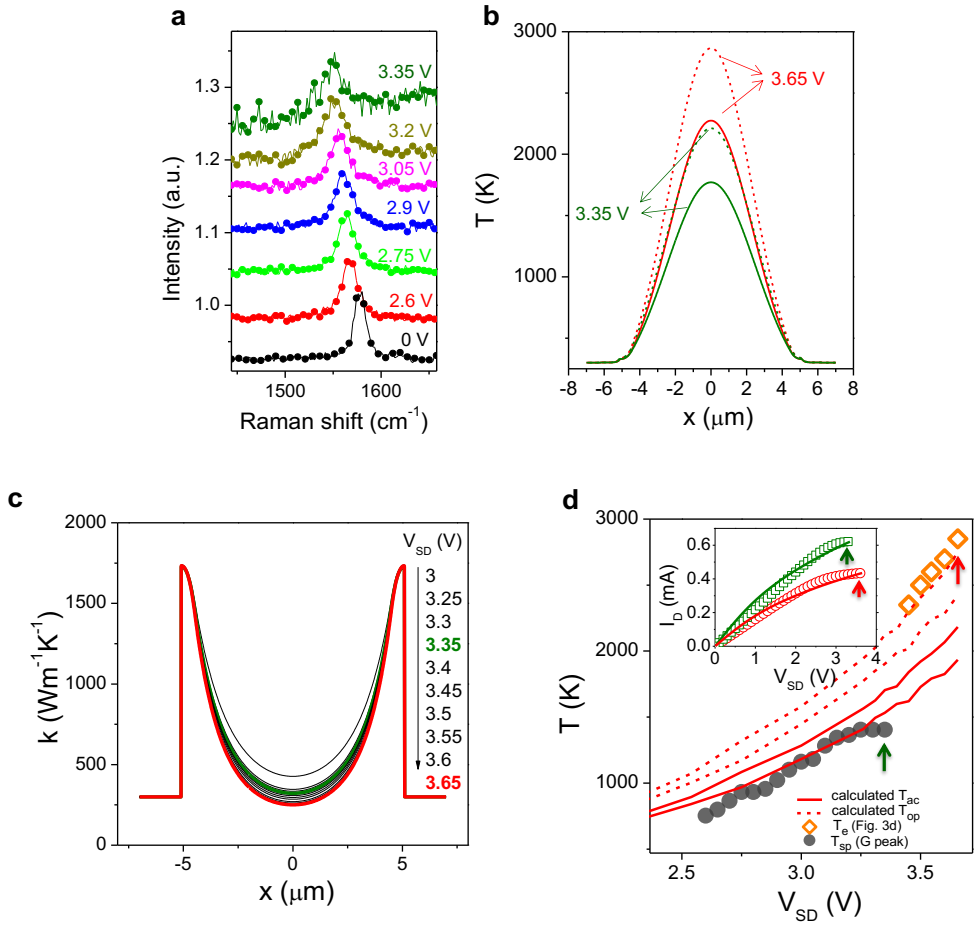

**Supplementary Fig. 19 | Intrinsic electro-thermal transport analysis in suspended ME tri-layer graphene a,** Raman spectra for various $V_{SD}$ of the tri-layer graphene. **b,** Temperature profiles along the suspended graphene length of APs (solid curves) and OPs (dashed curves) for $V_{SD}$ = 3.35 V and 3.65 V of the tri-layer graphene. **c,** Thermal-conductivity profile for various $V_{SD}$. **d,** Various temperatures as a function of $V_{SD}$ of the tri-layer graphene. Inset: $I_D$-$V_{SD}$ curves for radiation conditions with $V_{SD}$ = 3.35 V and 3.65 V indicated by green and red arrows, respectively (scattered points: experiments, solid and dashed curves: calculation results for $T_{ac}$ and $T_{op}$, respectively).

## B. Monolayer graphene

| $V_{SD}$ (V) | $\mu$ (cm$^2$V$^{-1}$s$^{-1}$) | $W$ (μm) | $T_{op}$ (K) | $T_{ap}$ (K) |
|---|---|---|---|---|
| 2.7 | 10000, 10250 | 0.784, 0.765 | 2634, 3039 | 1979, 2270 |
| 2.6 | 10000, 11500 | 0.796, 0.705 | 1802, 3016 | 1380, 2254 |
| 2.5 | 10000, 12700 | 0.87, 0.705 | 1381, 2951 | 1077, 2200 |
| 2.3 | 10000, 12700 | 1.15, 0.92 | 975, 1474 | 785, 1144 |
| 2 | 10000, 12700 | 1.63, 1.28 | 665, 838 | 562, 687 |
| 1.6 | 10000, 12700 | 1.93, 1.52 | 471, 525 | 423, 462 |

**Supplementary Table 2.** Parameters used for electro-thermal simulations for the monolayer graphene

We consider $\mu = 10000$ cm$^2$V$^{-1}$s$^{-1}$ as a minimum mobility for the suspended monolayer graphene[14]. The minimum mobility provides a lower bound of $T_{op}$. Recent experiments with the Raman spectroscopy have shown that the thermal conductivity of suspended monolayer graphene ranges from 2000 to 3000 Wm$^{-1}$K$^{-1}$ at room temperature[15-17]. During searching a proper thermal conductivity, $\kappa = 2700$ Wm$^{-1}$K$^{-1}$ was chosen through entire calculations. We also fixed $\gamma$ and $\beta$ as 1.92 and 1.7 as the tri-layer graphene case. Supplementary Table 2 shows parameters of $\mu$, $W$ and resultant $T_{op}$ and $T_{ap}$ of lower and upper bounds. For $V_{SD} = 2.7$ V, the minimum value of $W$ was found near $T_{ap} = 2300$ K, which is regarded as a burning temperature, $T_b$. Here, a nonequilibrium OP coefficient $\alpha$ for the suspended monolayer graphene was found as 0.39 as the best-fit result. For $V_{SD} = 2.6$ V, $W = 0.705$ μm was found as the width just before burning. When we consider $W = 0.705$ μm as a possible minimum width, we get $\mu(T_0) = 12700$ cm$^2$V$^{-1}$s$^{-1}$ as a possible maximum mobility for the examined graphene for $V_{SD} \leq 2.5$ V, which gives upper bounds of $T_{op}$ for $V_{SD} \leq 2.5$ V.

## C. Small grain size few-layer CVD graphene

For the suspended CVD few-layer graphene cases ($n \sim 3$) with a trench depth, $D = 300$ nm, we only observe monotonic thermal emission spectra for $L = 3$ μm and 4 μm CVD graphenes as shown in Supplementary Fig. 20a. Signal to noise ratio for the spectra with E > 2 eV is very low. Supplementary Fig. 20c shows $I_D$-$V_{SD}$ curves (scattered points) of the $L = 4$ μm CVD graphene, where the emission spectrum is obtained at $V_{SD} = 7.1$ V indicated by an arrow. By fitting with a grey-body theory[18,19], we get $T_e$ for $L = 4$ μm and 3 μm CVD graphenes of ~ 1300 K and ~ 1200 K, respectively, as indicated by dashed curves in Supplementary Fig. 20a. In addition, calculation results considering the interference effect of the thermal emissions with $D = 300$ nm do not show apparent spectral modulations for $T_e \sim 1300$ K and 1200 K as shown in Supplementary Fig. 20b (also see Supplementary Fig. 10d). Those are also consistent with the experimental spectra. Interestingly, in the electro-thermal model, we find that the best result as shown in Supplementary Fig. 20c is obtained for α = 0 (i.e., $T_{ap} = T_{op}$) with $T_{ap} \sim 1300$ K at the centre of the channel (see Supplementary Fig. 20d), which is consistent with $T_e$ obtained by the thermal emission model (with $W = 2$ μm, $\mu = 900$ cm$^2$V$^{-1}$s$^{-1}$, $\kappa = 2200$ Wm$^{-1}$K$^{-1}$, $\beta = 0.3$, $\gamma = 0.25$). For small grain size CVD few-layer graphene grown on a SiO$_2$/Si substrate by a plasma-assisted CVD process[1], it has been found that the size of grain boundary is ~ 20 nm as shown in Supplementary Fig. 4a. In this case, charge carriers could experience frequent scattering events by grain boundaries before emission of OPs, considering intrinsic graphene OP activation length is ~ 200 nm[8]. This could result in no enhanced OP population and $T_{ap} = T_{op}$ for up to $F \sim 1.7$ V/μm contrary to the suspended ME graphenes.

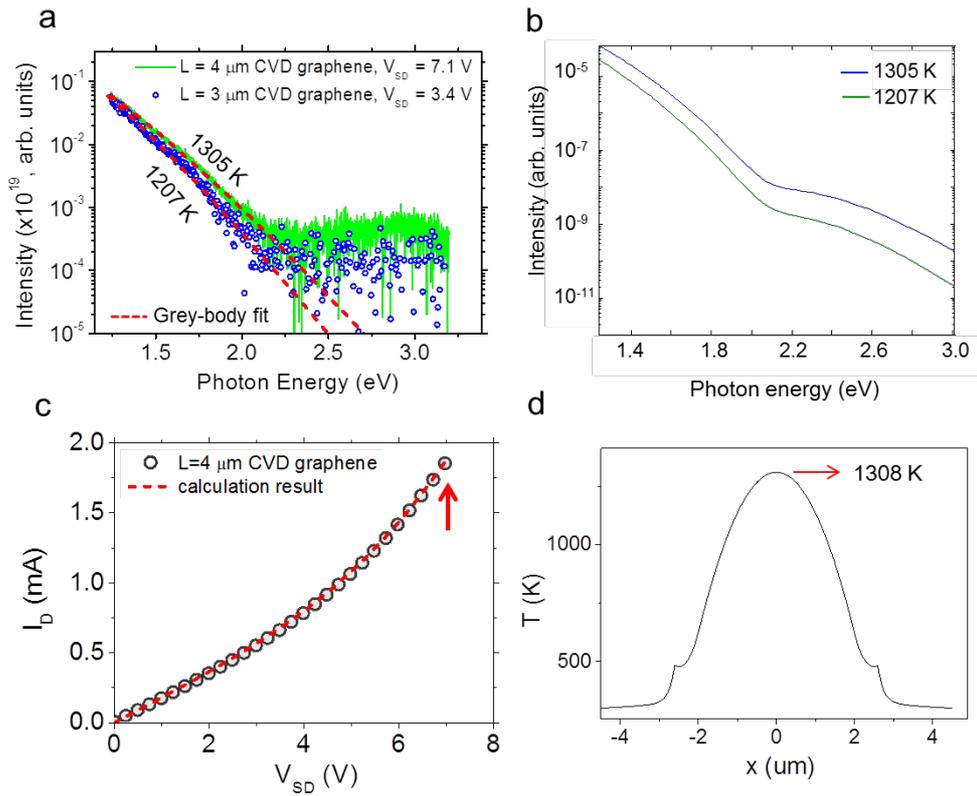

**Supplementary Fig. 20 | Suspended CVD few-layer graphene: light emission spectra and electro-thermo model**. **a,** Thermal emission spectra for $L$ = 4 μm and 3 μm CVD graphenes with $D$ = 300 nm. Dashed curves: grey-body fit results. **b,** Calculated spectra based on the interference effect of thermal emissions with $D$ = 300 nm at two different temperatures. **c,** Electrical transport ($I_D$ -$V_{SD}$ curve) of $L$ = 4 μm CVD few-layer graphene (scattered points: experimental data, dashed curve: calculation result based on the electro-thermal model). **d,** AP temperature profile along the $L$ = 4 μm graphene channel, where the maximum temperature (at the centre of the channel) is ~ 1300 K.

## S9. Thermal radiation power of electrically biased suspended ME graphene

In suspended ME monolayer graphene, we have observed bright visible light emission. To estimate the energy dissipation through radiation across all wavelengths, we calculate the radiated power given by the Stefan-Boltzmann law, $J = \epsilon \sigma T^4$, where $\epsilon$ is the emissivity of graphene (0.023 for monolayer graphene), $\sigma$ is Stefan's constant ($5.670 \times 10^{-8}$ W/m$^2$K$^4$) and $T$ is the electron temperature ($T$ is measured based on the carefully calibrated spectrometer). From the simulation results of light emission spectra (Fig. 2a) and intrinsic electro-thermo transport (Fig. 4c) from suspended ME monolayer graphene, we can estimate the electron temperature is about ~ 2800 K at applied electrical power $P_e = 1.756 \times 10^7$ W/m$^2$ ($V_{SD} = 2.7$ V, $I_D = 125$ μA). At the estimated electron temperature of ~2800 K, the radiated power ($P_r$) by the Stefan-Boltzmann law is $8.015 \times 10^4$ W/m$^2$, so energy dissipation through radiation ($P_r/P_e$) is ~ $4.45 \times 10^{-3}$. This remarkable light emission efficiency from suspended graphene is 1000 fold enhanced compare to the unsuspended graphene[19] (~$10^{-6}$). The efficient thermal radiation in suspended graphene is due to the elimination heat flow to the substrate, which is the dominant contributor to energy dissipation in unsuspended structures.

We also compare the radiation power efficiency of suspended ME multi-layer graphene, which depends on the emissivity of multi-layer graphene (n ~ 120). When we assume the emissivity of multi-layer graphene as 0.5 ~ 0.9[20], the radiated power ($P_r$) by the Stefan-Boltzmann law is 0.86 ~ $1.55 \times 10^4$ W/m$^2$ (T ~ 2350 K estimated from spectrum data) and applied electrical power $P_e$ ~ $1.71 \times 10^9$ W/m$^2$ ($V_{SD} = 4.9$ V, $I_D = 9.46$ mA), so energy dissipation through radiation ($P_r/P_e$) is 0.5 ~ $0.9 \times 10^{-3}$. Significant decrease in thermal radiation efficiency of suspended multi-layer graphene, compared to the monolayer graphene, could be due to the increase in the heat dissipation along graphene channel and the smallness of the difference between $T_{ap}$ and $T_{op}$ in multi-layer graphene.

Direct comparison between the efficiency of atomically thin graphene light emitter and that traditional bulk incandescent lamp is not trivial, because radiation efficiency depends on the operation electrical power. For example, luminous efficiency (at 400 ~ 700 nm wavelength) of typical tungsten light bulbs decrease as the electrical power is reduced (from 2% at 100 W, to 0.7% at 5 W). Thus, from a naive extrapolation of the known efficiency vs. power relation of typical tungsten incandescent lamps, we roughly estimate that the luminous efficiency of incandescent lamps (featureless black-body radiation) is 0.032% at 340 µW, which is lower than that of our suspended monolayer graphene light emitter due to enhancement of radiation spectrum within 400 ~ 700 nm by strong interference effect. Furthermore, traditional incandescent lamps do not operate at very low electrical power (~ 400 µW) due large heat capacitance of bulk materials and large heat dissipation. For these reasons, our graphene visible light emitter will play an important role where traditional incandescent technology is not applicable.

# S10. SEM image of mechanical failure of suspended graphene after remarkably bright visible light emission under a high electric field

After observation of remarkably bright visible light emission from suspended graphene under a high electric field (> 1 V/μm), suspended graphene devices are broken. At the local hot spot in the vicinity of the centre of suspended graphene approaching an extremely high temperature ($T_e$ > 3000 K), carbon atoms are sublimed and, in turn, thermally-generated defects are propagated. Most of such failures in our suspended graphene devices occur at the centre of the devices, where the temperature is highest as shown in Fig. 4c.

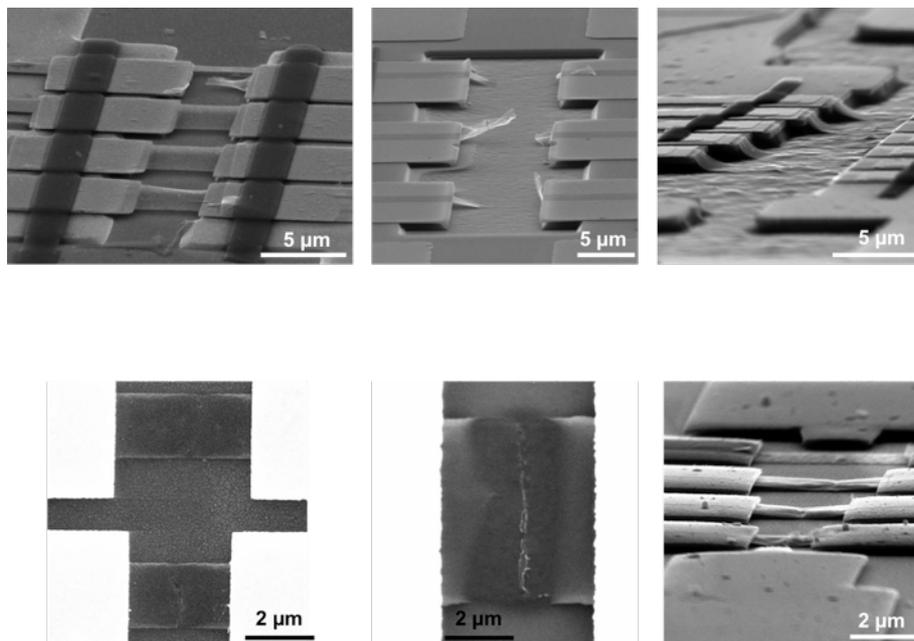

**Supplementary Fig. 21 | High-current/temperature-induced mechanical failure of suspended graphene after extremely bright visible-light emission.** SEM image of suspended (upper) ME graphene and (lower) CVD graphene.

# References for Supplementary Information